\documentclass[reprint,amsmath,amssymb,aps,pra,]{revtex4-2}

\usepackage{tikz}
\usetikzlibrary{matrix,arrows}
\usepackage{graphicx}
\usepackage{dcolumn}
\usepackage{bm}
\usepackage{bbold}
\usepackage{mathtools} 
\usepackage{dsfont} 
\usepackage{hyperref}
\hypersetup{
 colorlinks   = true,
 linkcolor = blue,
 citecolor    = blue,
 urlcolor = blue
}

\usepackage{balance}
\usepackage{lastpage}

\usepackage{soul}

\begin{document}

\preprint{APS/123-QED}

\title{Nonadiabatic coupled-qubit Otto cycle with bidirectional operation and efficiency gains}

\author{Cleverson Cherubim}
 \email{cleverson\_cherubim@id.uff.br}
\author{Thiago R. de Oliveira}%
 \email{troliveira@id.uff.br}
  \author{Daniel Jonathan}%
 \email{djonathan@id.uff.br}
\affiliation{Instituto de F\'isica, Universidade Federal Fluminense, Av. Gal. Milton Tavares de Souza s/n, Gragoat\'a 24210-346, Niter\'oi, RJ, Brazil}%
%



\date{\today}

\begin{abstract}
We study a quantum Otto cycle that uses a 2-qubit working substance whose Hamiltonian does not commute with itself at different times during unitary strokes. We investigate how the cycle responds to the loss of quantum adiabaticity when these strokes are operated with a finite duration. We find that qualitative features such as the possibility of counter-rotating cycles operating as heat engines, or a cycle efficiency that can increase with a decrease in the temperature difference between the baths, are resilient even to highly nonadiabatic strokes. However, cycle efficiency rapidly decreases, although it can still remain above the standard Otto value for small degrees of quantum nonadiabaticity.
\end{abstract}


\maketitle

\section{\label{sec:level1}Introduction}
Heat engines have been present in the study of thermodynamics \cite{callen1985thermodynamics,111837181X} since its very beginning. This is not only due to their practical applications, but because they present a useful theoretical framework to describe the limitations imposed by the second law of thermodynamics on the conversion of heat to work. 
More recently, heat engines have also been at the forefront of quantum thermodynamics \cite{10.1088/2053-2571/ab21c6,binder2019thermodynamics}, which deals with the limitations imposed by the laws of thermodynamics on small systems subjected to both thermal and quantum fluctuations.  One of the most discussed questions in this scenario is whether heat engines may exploit quantum properties in order to obtain some advantage over, or operate differently from, their classical counterparts \cite{PhysRevX.7.031044, Abah_2014, PhysRevE.86.051105, PhysRevLett.112.030602, 2015NatSR512953H, Niedenzu2016, PhysRevE.93.052120, PhysRevB.96.104304, PhysRevE.90.012119, PhysRevE.88.032103, deoliveira2020efficiency, PhysRevLett.127.190604}.

Recently, two of us  have pointed out \cite{deoliveira2020efficiency} that such advantages and differences can occur in the case of a two-heat-bath quantum Otto cycle \cite{doi:10.1063/1.461951, PhysRevLett.93.140403, PhysRevE.76.031105, PhysRevE.83.031135}. Before describing these, it is worth clarifying for the general reader that in an (ideal) `quantum'  Otto cycle not only is the working substance described quantum-mechanically, but the work strokes are `adiabatic' in the quantum sense, i.e.,  unitary evolutions representing a slow time-dependent change in a parameter of the Hamiltonian, during which there is no change in the energy occupation probabilities, while the energy eigenvalues and eigenstates evolve smoothly. This is not always the same as an adiabatic process in the usual thermodynamic sense, which is a quasi-static process where the system remains always infinitesimally close to thermal equilibrium. In the quantum adiabatic case the system may be driven far from the equilibrium thermal state of its time-evolving Hamiltonian \cite{PhysRevE.72.056110}. In this scenario, it was shown in Ref. \cite{deoliveira2020efficiency} that if some levels of the  working substance are ‘idle’, in the sense that they do not couple to the external work source or sink during the adiabatic strokes, while the remaining levels shift proportionately to the adiabatic control parameter, then it is possible to obtain enhanced efficiency relative to the standard Otto value. In these cases, the gain can be interpreted  as due to heat partially propagating backwards, from the cold to the hot bath, via the idle levels. A similar mechanism allows `counter-rotating' engine cycles that run in reverse to their ordinary sense but still output work, requiring only an exchange in the roles of the hot and cold baths. Other unusual phenomena were also reported to occur in such engines, such as situations where the engine efficiency increased while the bath temperature difference decreased.

However, Ref. \cite{deoliveira2020efficiency} studied an idealized scenario where no explicit time dependence was taken into account. Both the thermalization and unitary adiabatic strokes of the quantum Otto cycle were assumed for the most part to be perfect, something which would require infinite time. Some weaker conditions were also considered: for example, it was argued that, by continuity, the qualitative results mentioned above would be robust to small imperfections, such as almost-complete thermalization or almost-perfect adiabatic strokes. In addition, a full example was worked out in which the system Hamiltonian $\hat{\rm{H}}(\rm{t})$ commuted with its own time evolution throughout the `adiabatic' strokes - in which case these strokes could in fact be realized at any desired speed, outside the quantum adiabatic regime.

In this article, we propose to extend this investigation to similar cycles where this special condition is not satisfied, i.e., when $[\hat{\rm{H}}(\rm{t}), \hat{\rm{H}}(\rm{t'})] \neq 0$ at different times during the unitary strokes \cite{PhysRevB.101.054513,PhysRevB.94.184503,Cakmak2016,PhysRevE.99.032108}. We would like to know, for example, whether the engine properties mentioned above can endure the lack of quantum adiabaticity that occurs when these strokes are completed in finite time. As we will see, this is indeed the case, although they degrade as the system decreases its quantum adiabaticity. For example, the temperature ranges where the cycle operates as a regular or `counter-rotating' engine reduce in scope, but do not vanish. In addition, their efficiency is systematically reduced, until gains relative to the standard Otto efficiency are no longer possible. Finally, we find that there is no longer a direct connection between the presence of a reversed heat flux and an increase in efficiency - it is possible to have the former without the latter.

\section{\label{description}Description of the system}

The working substance model we propose to use to investigate these questions is the following: a system of 2 spin-1/2 particles interacting via an XY-type coupling, and subjected to an external classical field $\rm{h}(\rm{t})\hat{z}$ whose amplitude follows some externally determined time-dependence. The source of this field acts as the work sink for this heat engine. The Hamiltonian is
\begin{equation}
\label{ws_hamiltonian}
\hat{\rm{H}}(\rm{t})=\rm{J}_x\hat{\sigma}_{1x}\hat{\sigma}_{2x}+\rm{J}_y\hat{\sigma}_{1y}\hat{\sigma}_{2y}+h(t)[\hat{\sigma}_{1z}+\hat{\sigma}_{2z}],
\end{equation}
with $\hat{\sigma}_{ix}$, $\hat{\sigma}_{iy}$ and $\hat{\sigma}_{iz}$ being the Pauli matrices associated to the $i_{\rm{th}}$ qubit. $\rm{J}_x$, $\rm{J}_y$ represent respectively the coupling strengths along the $x$ and $y$ directions \footnote{Note that, throughout this article, we use units such that $\hbar$, $k_B$ and the magnetic moment $\mu$ of each spin are all equal to 1}. As mentioned above, this choice is motivated by the fact that, unlike the isotropic Heisenberg-type interaction studied in \cite{deoliveira2020efficiency}, here $\hat{\rm{H}}(\rm{t})$ does not commute with itself at different times.

This Hamiltonian can be easily solved analytically (see Appendix \ref{Apdx-A}). Its eigenenergies are given by: $\varepsilon_4~=~\sqrt{4\rm{h(t)}^2+(\rm{J}_x-\rm{J}_y)^2}$, $\varepsilon_3= (\rm{J}_x+\rm{J}_y)$, $\varepsilon_2=-\varepsilon_3$ and $\varepsilon_1=-\varepsilon_4$. They can be divided into two types: those dependent on the `work parameter' $\rm{h(t)}$, namely $\varepsilon_1$ and $\varepsilon_4$, and those independent of it, namely $\varepsilon_2$ and $\varepsilon_3$. These two classes will play a fundamental role in the behavior of the cycle, as explained below. The corresponding eigenstates also form two families: $|\varepsilon_1\rm(t)\rangle$ and  $|\varepsilon_4(\rm{t})\rangle$ are real, time-dependent combinations of the states $|\uparrow \uparrow \rangle$ and $|\downarrow \downarrow \rangle$:
\begin{align} 
|\varepsilon_4\rm(t)\rangle =& \alpha_{+}(t) |\uparrow \uparrow \rangle + \alpha_{-}(t)|\downarrow \downarrow \rangle, \nonumber \\
|\varepsilon_1\rm(t)\rangle =& \alpha_{-}(t)|\uparrow \uparrow \rangle  -\alpha_{+}(t)|\downarrow \downarrow \rangle \label{epsilon1} ,
\end{align}
where 
\begin{align}\label{alphapm}
\alpha_{\pm}(t) = \sqrt{\frac{1}{2}\left(1\pm\frac{2\rm{h(t)}}{\sqrt{4\rm{h(t)}^2+(\rm{J}_x-\rm{J}_y)^2}}\right)},
\end{align}
while  $|\varepsilon_2\rangle$ and 
$|\varepsilon_3\rangle$ are the Bell states $\left(|\uparrow \downarrow \rangle \mp |\downarrow \uparrow \rangle\right)/\sqrt{2}$, independently of $\rm{h(t)}, \rm{J}_x$ or $\rm{J}_y$.

The cycle is implemented following an Otto-type recipe, with two thermalization strokes and two time dependent strokes. In the latter, the external field $\rm{h}(\rm{t})$ is varied back or forth between two extreme values $\rm{h}_1$ and $\rm{h}_2<\rm{h}_1$ during a finite time interval of length $\tau$, while the system is kept out of contact with the heat reservoirs. 

The unitary evolution during these strokes is nontrivial, since $[\hat{\rm{H}}(\rm{t}), \hat{\rm{H}}(\rm{t'})] \neq 0$. It is also important to notice that this evolution will generally take the working substance from a thermal state to an out-of-equilibrium state. 

The complete cycle is detailed in the following set of instructions:

\textit{First stroke: }The working substance starts in a thermal state $\hat{\varrho}_1=e^{-\beta_1\hat{\rm{H}}_1}/\rm{Z}_1$ at temperature $\rm{T}_1=\beta_1^{-1}$ $(k_{\rm{B}}=1)$, with $\hat{\rm{H}}_1=\hat{\rm{H}}(0)$ and $\rm{Z}_1=\rm{Tr}(e^{-\beta_1\hat{\rm{H}}_1})$. The external field is then varied according to some specific time-dependence $\rm{h(t)}$, during $0\leq t \leq \tau$. The spins undergo the corresponding quantum non-adiabatic unitary transformation $\hat{\rm{U}}(\tau)=\mathcal{T}e^{-i \int_0^{\tau}\rm{dt}\hat{\rm{H}}(\rm{t})}$, $\mathcal{T}$ being the time-ordering operator, finishing the stroke in the state $\hat{\varrho}_2=\hat{\rm{U}}(\tau)\hat{\varrho}_1\hat{\rm{U}}^{\dagger}(\tau)$. Adopting the usual quantum-mechanical definition of work as the (average) energy exchange due to external changes in a system's Hamiltonian \cite{0305-4470-12-5-007, doi:10.1063/1.446862}, then only work is exchanged in this stroke between the working substance and the magnetic field. It may be calculated via the change in internal energy, $\rm{W}_{1\to2}=\mathbb{E}_2-\mathbb{E}_1$, with $\mathbb{E}_1=\rm{Tr}(\hat{\varrho}_1\hat{\rm{H}}_1)$ and $\mathbb{E}_2=\rm{Tr}(\hat{\varrho}_2\hat{\rm{H}}_2)$. As we show in section \ref{sec:workheat} below, this work is always positive, i.e. this stroke is analogous to the compression stroke of an ideal-gas based Otto engine.  

It is worth noting that there is an ongoing debate on how to define work in situations involving time-dependent eigenvectors or states displaying energy-basis coherence \cite{Allahverdyan_2004, RevModPhys.83.771, Niedenzu2018, PhysRevE.102.062152, 2019arXiv191201939A, Su_2018,2019arXiv191201983A}. It is also possible to consider work as a stochastic variable, defined as the difference between energy measurements at the beginning and end of a stroke (see e.g. Sec. IIIA of \cite{RevModPhys.83.771}). From this one can obtain an average work, which in some cases may differ from the difference between average energies that we have used above. One can also consider the `ergotropy' of the state  \cite{Allahverdyan_2004, Niedenzu2018}, the maximum energy that can be extracted from it via a cyclical unitary transformation. Since this process does not change the von Neumann entropy, this energy can arguably be considered work; 
We shall not discuss these alternate definitions in this paper.

\textit{Second stroke: }In this stroke the working substance is put in contact with a heat bath at temperature $\rm{T}_2=\beta_2^{-1}$ and allowed to relax with a fixed Hamiltonian $\hat{\rm{H}}_2=\hat{\rm{H}}(\tau)$ until it reaches the thermal state $\hat{\varrho}_3=e^{-\beta_2\hat{\rm{H}}_2}/\rm{Z}_2$ with $\rm{Z}_2=\rm{Tr}(e^{-\beta_2\hat{\rm{H}}_2})$. Since the external  parameters of the system remain fixed, by definition only heat is exchanged with the bath, given by the variation in average internal energy, $\rm{Q}_2=\mathbb{E}_3-\mathbb{E}_2$, with $\mathbb{E}_3=\rm{Tr}(\hat{\varrho}_3\hat{\rm{H}}_2)$.

 \begin{figure}[t]
		\includegraphics[scale=.5]{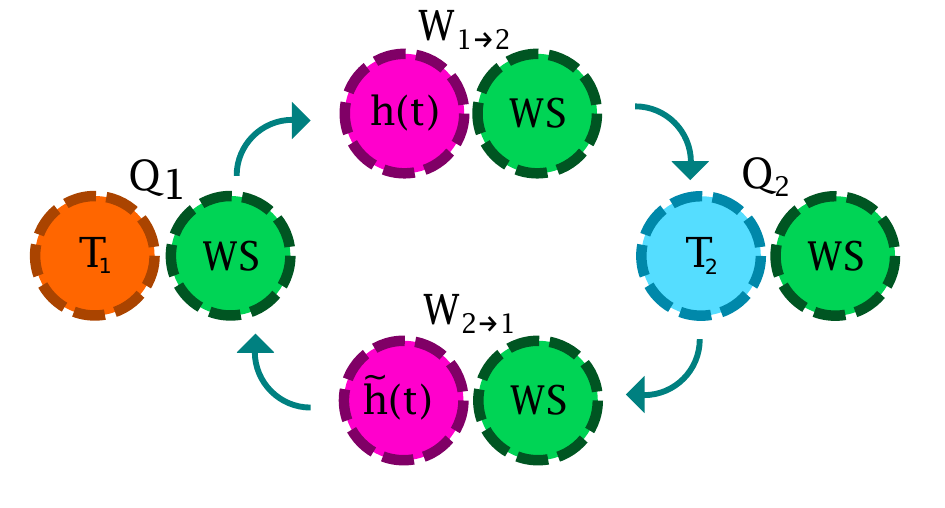}
		\centering
		\caption{\label{cycle_sketch} Sketch of the cycle, with its four strokes depicted. Clockwise from the top:  first the working substance (WS) exchanges work $\rm{W}_{1\to2}$ as the magnetic field $\rm{h}(t)$ is varied from $\rm{h}_1$ to $\rm{h}_2$ during an interval of finite length $\tau$. Next it thermalizes with the reservoir at  $\rm{T} = \rm{T}_2$, exchanging heat $\rm{Q}_2$. It then exchanges work $\rm{W}_{2\to1}$ as the field is reversed back from $\rm{h}_2$ to $\rm{h}_1$, following $\tilde{\rm{h}}(t) = \rm{h}(\tau - t)$ . Finally the WS thermalizes with the reservoir at  $\rm{T} = \rm{T}_1$, exchanging heat $\rm{Q}_1$ and returning to its initial state. Note that $\rm{T_{2}}$ is not necessarily higher than $\rm{T_{1}}$.}
	\end{figure}

\textit{Third stroke: }The system is removed from contact with the heat bath, and the external field is reversed back to its initial value. We assume the time-dependence $\rm{\tilde{h}(t)}$ during this reversal follows the rule $\rm{\tilde{h}(t)} = \rm{h(\tau-t)}$. The spins undergo the corresponding reversed unitary operation $\hat{\rm{V}}(\tau)=\mathcal{T}e^{-i \int_{0}^{\tau}\rm{dt}\hat{\rm{H}}'(\rm{t})}$, with $\hat{\rm{H}}'(\rm{t})=\hat{\rm{H}}(\rm{\tau-\rm{t}})$, generating the state $\hat{\varrho}_4=\hat{\rm{V}}(\tau)\hat{\varrho}_3\hat{\rm{V}}^{\dagger}(\tau)$. 
The work exchanged in this stroke is given by $\rm{W}_{2\to1}=\mathbb{E}_4-\mathbb{E}_3$, with $\mathbb{E}_3=\rm{Tr}(\hat{\varrho}_3\hat{\rm{H}}_2)$ and $\mathbb{E}_4=\rm{Tr}(\hat{\varrho}_4\hat{\rm{H}}_1)$. As we show in section \ref{sec:workheat} below, this work is generally negative, i.e. this stroke is analogous to the expansion stroke of an ideal-gas based Otto engine.

It should be noted that $\hat{\rm{U}}(\tau)$ and $\hat{\rm{V}}(\tau)$ are not independent. In general, it can be shown \cite{PhysRevB.101.054513} that, if the Hamiltonian has the property that $\hat{\rm{K}}\hat{\rm{H}}(\rm{t})\hat{\rm{K}}^{\dagger} = \hat{\rm{H}}(\rm{t})$ for some antiunitary operator $\hat{\rm{K}}$ and all \rm{t}, then
\begin{equation}\label{reverseevolution}
  \hat{\rm{V}}(\tau) = \hat{\rm{K}} \hat{\rm{U}}^{\dagger}(\tau) \hat{\rm{K}}^{\dagger}. 
\end{equation}
This is indeed the case here, since $\hat{\rm{H}}(\rm{t})$ is represented by a real matrix in the $\hat{\sigma}_{1z}\hat{\sigma}_{2z}$ basis. Eq. (\ref{reverseevolution}) therefore holds if $\hat{\rm{K}}$ is the complex conjugation operator for this basis. (Note Eq. (\ref{reverseevolution}) does \emph{not} hold if $\hat{\rm{K}}$ is the standard time-reversal operator, since this operator flips the individual $\hat{\sigma}_z^{j}$ terms \cite{Messiah62}).

\textit{Fourth stroke: } The final stroke is composed of another relaxation process, this time with the thermal bath at temperature $\rm{T}_1=\beta_1^{-1}$, while the Hamiltonian is kept constant and equal to $\hat{\rm{H}}_1=\hat{\rm{H}}(0)$. The system finishes the cycle returning to the initial state $\hat{\varrho}_1=e^{-\beta_1\hat{\rm{H}}_1}/\rm{Z}_1$. In analogy with the second stroke, the heat exchanged with the bath at $\rm{T} = \rm{T}_{1}$ is $\rm{Q}_1=\mathbb{E}_1-\mathbb{E}_4$. The full protocol is sketched in Fig. \ref{cycle_sketch}.

In what follows, we denote the net work \emph{injected into} the system along the entire cycle as 
\begin{equation}\label{Wconvention}
\rm{W}_{\rm{cyc}}= \rm{W}_{1\to2}+\rm{W}_{2\to1} =-( \rm{Q}_1+\rm{Q}_2).
\end{equation}

Thus, $\rm{W}_{\rm{cyc}}<0$ corresponds to a heat engine cycle. For $\rm{W}_{\rm{cyc}}>0$, the cycle can operate in different regimes, as we will discuss in Sec. \ref{regimes}. 

\subsection{\label{non_adiabatic}Quantum non-adiabatic evolution}

The quantum adiabatic regime, where a system remains in an energy eigenstate even as its energy slowly changes, holds when an externally controlled parameter in its Hamiltonian is changed at a rate that is much smaller than the smallest frequency gap between energy levels. Conversely, executing a unitary stroke such as $\hat{\rm{U}}(\tau)$ in a sufficiently short interval $\tau$ will generally lead to non-negligible transition probabilities between energy eigenstates that are coupled during the stroke. For example, in the present case, such transitions will be induced between levels $|\varepsilon_1\rangle$ and $|\varepsilon_4 \rangle$. Such transitions are considered in the literature to represent `quantum internal friction' \cite{Rezek_2006}. They can be seen as arising due to the difference between the internal time scales, governed by the working substance's energy gaps, and the time scales of the external driving field \cite{e12081885}.

It turns out that, because of the symmetry expressed in Eq. (\ref{reverseevolution}), the forward and reverse unitary strokes $\hat{\rm{U}}(\tau)$ and $\hat{\rm{V}}(\tau)$ are \emph{microreversible}, in the sense that the transition probability between two energy eigenstates $|\varepsilon_j^{(1)}\rangle=|\varepsilon_j(0)\rangle$ and $|\varepsilon_i^{(2)}\rangle=|\varepsilon_i(\tau)\rangle$ along the first stroke, $|\langle\varepsilon_i^{(2)}|\rm{\hat{U}}(\tau)|\varepsilon_j^{(1)}\rangle|^2$, equals the probability of the reversed transition in the third stroke, $|\langle\varepsilon_j^{(1)}|\hat{\rm{V}}(\tau)|\varepsilon_i^{(2)}\rangle|^2$. To see this, note that,  $|\varepsilon_i^{(k)}\rangle$ have \emph{real} coefficients in the $\hat{\sigma}_{1z}\hat{\sigma}_{2z}$  basis, so that $\hat{\rm{K}} |\varepsilon_i^{(k)}\rangle = |\varepsilon_i^{(k)}\rangle $. It follows that \cite{PhysRevB.101.054513}
\begin{align}
\nonumber |\langle\varepsilon_j^{(1)}|\hat{\rm{V}}(\tau)|\varepsilon_i^{(2)}\rangle|^2 &=  
|(\langle\varepsilon_j^{(1)}| \hat{\rm{K}}^{\dagger})  \hat{\rm{V}}(\tau)  (\hat{\rm{K}} |\varepsilon_i^{(2)}\rangle) |^2 \\
\nonumber  &=|(\langle\varepsilon_j^{(1)}| \hat{\rm{K}}^{\dagger}) (\hat{\rm{K}} \hat{\rm{U}}^{\dagger}(\tau) \hat{\rm{K}}^{\dagger} ) (\hat{\rm{K}} |\varepsilon_i^{(2)}\rangle) |^2 \\
 \nonumber  &= |\langle\varepsilon_j^{(1)}| (\hat{\rm{K}}^{\dagger} \hat{\rm{K}} \hat{\rm{U}}^{\dagger}(\tau) \hat{\rm{K}}^{\dagger} \hat{\rm{K}}) |\varepsilon_i^{(2)}\rangle^{*} |^2 \\
\nonumber & = |\langle\varepsilon_j^{(1)}| \hat{\rm{U}}^{\dagger}(\tau) |\varepsilon_i^{(2)}\rangle^{*} |^2  \\
&=  |\langle\varepsilon_i^{(2)}| \hat{\rm{U}}(\tau) |\varepsilon_j^{(1)}\rangle|^2  \label{reverseprobs}
\end{align}
where we have used Eq. (\ref{reverseevolution})  and the facts that, for antiunitary operators, $\big(\langle a| \hat{\rm{K}} \big) |b\rangle = \left[ \langle a| \big(\hat{\rm{K}} |b\rangle\big) \right]^{*}$ and $\hat{\rm{K}}^{\dagger}\hat{\rm{K}} =  \hat{\mathds{1}}$ \cite{Messiah62}. It is therefore sufficient to calculate the transition probabilities only for the first unitary stroke.

In order to do this, it is convenient to expand the evolved states in the energy eigenbasis at time $\tau$, $\rm{U}(\tau)|\varepsilon_m(\rm{0})\rangle =\sum_{n}c_n^{(m)}(\tau)e^{i\theta_n(\tau)}|\varepsilon_n(\tau)\rangle$, with $\theta_n(\tau) \coloneqq -\int_0^{\tau}\varepsilon_n(t)dt$ being the dynamical phase and with the initial condition $c_n^{(m)}(0)=\delta_{nm}$. Replacing this expansion in the time-dependent Schr\"odinger equation, we obtain a set of ordinary differential equations for the energy amplitudes $c_n^{(m)}(t)$:

 \begin{equation}\label{amplitude-eq}
    	\begin{split}
    	\dot{c}_1^{(m)}\rm(t)&=\dot{\rm{h}}(\rm{t})\frac{e^{-2i\theta_1(\rm{t})}}{\varepsilon_1^2(t)}\left|\rm{J}_x-\rm{J}_y\right|c_4^{(m)}(t)\\
    	\dot{c}_2^{(m)}\rm(t)&=0\\
    	\dot{c}_3^{(m)}\rm(t)&=0\\
    	\dot{c}_4^{(m)}\rm(t)&=-\dot{\rm{h}}(\rm{t})\frac{e^{2i\theta_1(\rm{t})}}{\varepsilon_1^2(t)}\left|\rm{J}_x-\rm{J}_y\right|c_1^{(m)}(t)
    	\end{split}
	\end{equation}

As expected, $c_2^{(m)}\rm(t)$ and $c_3^{(m)}\rm(t)$ are time-independent, but $c_1^{(m)}\rm(t)$ and $c_4^{(m)}(\rm{t})$ are coupled, allowing for transitions between the corresponding eigenstates. Their values depend on the parameters $\rm{J}_x$, $\rm{J}_y$ but also on the specific function of time, or \emph{protocol} $\rm{h(t)}$ used to change the external field, as well as the total time $\tau$ taken to execute this protocol. This highlights the non-adiabaticity of the process. Note also that, since only levels $1$ and $4$ are coupled, we have in effect a two-level non-adiabatic evolution, similar to the one studied in  \cite{PhysRevB.101.054513}, but augmented by the two `idle' levels 2 and 3.

In a perfect adiabatic evolution, we would have $|c_{k}^{(k)}(\tau)|^2 \equiv 1$ for all $\tau$. Because the non-adiabatic coupling in Eq.(\ref{amplitude-eq}) generates coherence in the energy eigenbasis, this is no longer true for levels $k=1,4$. In fact, we can measure the deviation from perfect adiabaticity using the quantity
\begin{equation} \label{adiabaticity}
\rm{P}(\tau)\equiv |\langle\varepsilon_1^{(2)}|\rm{\hat{U}(\tau)}|\varepsilon_1^{(1)}\rangle|^2=|c_{1}^{(1)}(\tau)|^2.
\end{equation}
Note that, due to unitarity, $|c_{1}^{(4)}(\tau)|^2 = |c_{4}^{(1)}(\tau)|^2 = 1- \rm{P}(\tau)$, and  $|c_{4}^{(4)}(\tau)|^2 = \rm{P}(\tau)$ too. Thus, $\rm{P}(\tau)$  completely captures the degree of adiabaticity of the evolution \cite{PhysRevB.101.054513}.  The closer it gets to $1$, the more quantum-adiabatic the evolution is.

In general, we cannot determine an exact analytical solution for this quantity. An approximate solution, particularly appropriate for low degrees of nonadiabaticity, could in principle be obtained using adiabatic perturbation theory \cite{PhysRevA.78.052508,PhysRevE.100.032144}.
Here we opt instead to solve Eq. (\ref{amplitude-eq}) numerically, for a given choice of total stroke time $\tau$, coupling strengths $\rm{J}_x, \rm{J}_y$,  and protocol  $\rm{h(t)}$. Nevertheless, the limiting behaviors of $ \rm{P}(\tau)$  are simple to determine analytically, and are protocol-independent, apart from the endpoints $\rm{h}_{1,2}$. First of all, in the limit of very fast evolutions $(\tau \to 0)$ the system undergoes a `quench', i.e., the Hamiltonian is in effect changed instantaneously. In this case, as is well-known \cite{Messiah62} the state itself remains unchanged, i.e, $\rm{\hat{U}}(\tau) \to \hat{\mathds{1}}$. Using Eq. (\ref{epsilon1}), we thus get
\begin{align}
&\rm{P}(\tau \to 0) \to  P_{0}\equiv |\langle\varepsilon_1^{(2)}|\varepsilon_1^{(1)}\rangle|^2 \nonumber \\
 &= \frac{1}{2}\left[1+ \frac{4\rm{h_{1}} \rm{h_{2}} + (\rm{J_{x}} -\rm{J_{y}} )^{2}}{\sqrt{\left(4\rm{h_{1}^{2}} +(\rm{J_{x}} -\rm{J_{y}} )^{2} \right)\left(4\rm{h_{2}^{2}} +(\rm{J_{x}} -\rm{J_{y}} )^{2} \right)}}\right]. \label{Pinstant}
\end{align}

For the opposite limit of large $\tau$, it seems reasonable to assume  $\rm{P}(\tau) \to 1$, i.e., that the evolution will correspond to the (quantum) adiabatic limit.  This does indeed turn out to be so; however, we need to take some care. For general $\rm{\hat{U}}(\tau)$ this property is not always true, due to the possibility of level crossings. At these points, nonadiabatic transitions may potentially occur even for arbitrarily slow evolutions \cite{teufel2003adiabatic}. For our particular Hamiltonian, such crossings do happen between the eigenenergies $\varepsilon_1$ and $\varepsilon_2$  (and, simultaneously, between $\varepsilon_3$ and $\varepsilon_4$), whenever  $\rm{h(t)} = \sqrt{\rm{J}_x \rm{J}_y }$.
Nevertheless, this does not lead to nonadiabatic transitions, because the time evolution does not couple the eigenstates involved in the crossings, i.e., $\langle\varepsilon_2\rm(t)|\dot{\varepsilon}_1\rm(t)\rangle=\langle\varepsilon_3\rm(t)|\dot{\varepsilon}_4\rm(t)\rangle=0$. In this case, we may indeed apply the adiabatic approximation $\rm{P}(\tau) \to 1$ without worry  \cite{teufel2003adiabatic}. In other words, for large enough $\tau$, we can always expect the evolution to approach an ideal adiabatic stroke, like those studied in Ref. \cite{deoliveira2020efficiency}.

\begin{figure}[t]
		\includegraphics[scale=.55
		]{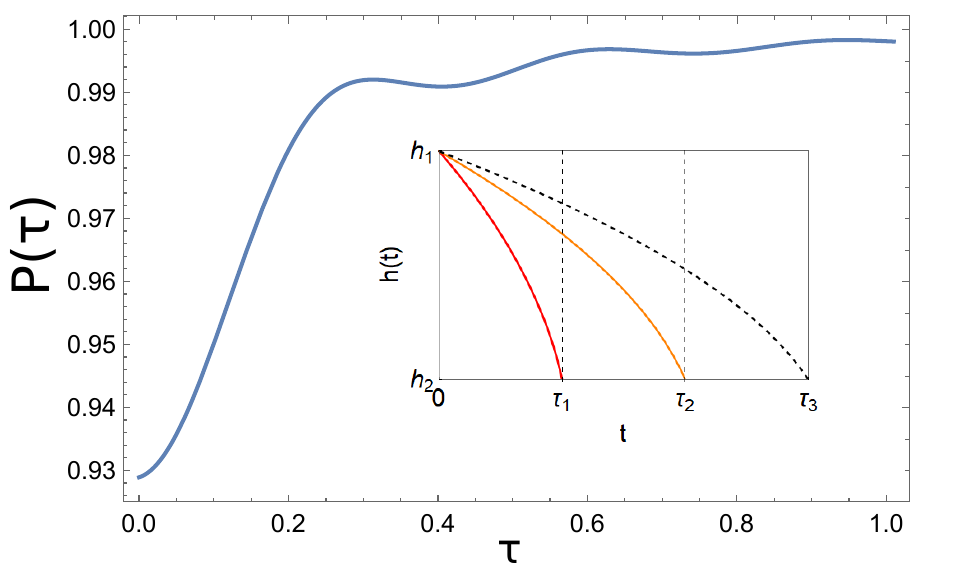}
		\centering
		\caption{\label{C_J1_J2} Adiabaticity coefficient $\rm{P}(\tau)$ (Eq.\,(\ref{adiabaticity})), as a function of stroke duration $\tau$, for the unitary evolution $\rm{\hat{U}}(\tau)$ 
with parameters $\rm{J}_x=10.0, \rm{J}_{y}=2.0$, $\rm{h}_{1}=4.0$, $\rm{h}_{2}=1.0$, and field $\rm{h(t)}$ varying according to the protocol in Eq.\,(\ref{protocol}).
As $\tau\to 0$,  the adiabaticity decreases towards the limit given in Eq.\,(\ref{Pinstant}): $\rm{P}(\tau\to 0) \to  |\langle\varepsilon_1^{(2)}|\varepsilon_1^{(1)}\rangle|^2\approx 0.929$. For $\tau\to\infty$, $\rm{P}(\tau)\to 1$ (adiabatic limit). Inset:  examples of  $\rm{h}(\rm{t})$ for different times of execution $\tau_1<\tau_2<\tau_3$. }
	\end{figure}

As mentioned in the Introduction, our main goal in this article is to examine the effects of nonadiabaticity on the phenomena described in \cite{deoliveira2020efficiency}.  We are thus particularly interested in parameter regimes where those phenomena occur. For example: when $\rm{J}_x=10.0, \rm{J}_y=2.0, \rm{h}_1=4.0$ and $\rm{h}_2=1.0$, it turns out the ideal adiabatic cycle operates as a counter-rotating engine, as will be discussed in the next section.
Fig.\,\ref{C_J1_J2} depicts the amount of non-adiabaticity that occurs in this case as a function of the stroke length $\tau$, for a particular protocol of the form
\begin{equation}\label{protocol}
\rm{h}(\rm{t})=\sqrt{\rm{h}_2^2  \left(t/\tau\right)+\rm{h}_1^2 \left(1-t/\tau\right)}.
\end{equation}

As expected, $\rm{P}(\tau)\approx 1$ for large $\tau$. In addition, as predicted by Eq.(\ref{Pinstant}),  $\rm{P}(\tau)\approx 0.929$ as $\tau\to 0$, i.e., in this case, even for a very fast stroke we still observe a relatively high degree of quantum adiabaticity. Note $\rm{P}(\tau)$ is not necessarily monotonic in $\tau$ (although, at least in this example, it attains its minimum value at $\tau \to 0$). In the next section, we discuss the effects non-adiabaticity can have on the properties of the thermodynamic cycle.

\section{\label{thermoadiab} Nonadiabatic Otto Cycle}

\subsection{Nonadiabatic Work and Heat Exchanges} \label{sec:workheat}

As might be expected, the presence of nonadiabatic transitions directly affects the work and heat exchanges during the cycle. In a nutshell, when compared to the corresponding adiabatic cycle, these transitions result in: more work being required during the first (`compression') stroke, less work being extracted in the third (`expansion') stroke, and more heat being dissipated into (or less  extracted out of) the thermal baths.

To see all this, let us write explicitly expressions for the average internal energies of the working substance at the end of the first and third strokes, as defined in Sec.~\ref{description}:
\begin{equation}
	\begin{split}
 \mathbb{E}_2&=\sum_{i,j=1}^{4}\varepsilon_i^{(2)}\rm{p}_{j}^{(1)}|c_{i}^{(j)}(\tau)|^2,\\
 \mathbb{E}_4&=\sum_{i,j=1}^{4}\varepsilon_j^{(1)}\rm{p}_{i}^{(2)}|c_{i}^{(j)}(\tau)|^2.
	\end{split}
	\end{equation}
Here $\varepsilon_i^{(k)}$  denote the eigenenergies of each endpoint Hamiltonian $\hat{\rm{H}}_k$, $\rm{p}_{j}^{(k)}=e^{-\beta_k\varepsilon_j^{(k)}}/\rm{Z}_k$ are the thermal populations at the respective inverse temperatures $\beta_{k}$, and in the second line we have used Eq. (\ref{reverseprobs}). Note also that, assuming full thermalization occurs in the second and fourth strokes, the respective average energies after each stroke, $\mathbb{E}_3$ and $ \mathbb{E}_1$, are insensitive to the nonadiabaticity. 

The work exchanged in each unitary stroke can then be expressed as a sum of adiabatic and non-adiabatic contributions, i.e. $\rm{W}_{a\to b } = \rm{W}_{a \to b }^{ad}+ \rm{W}_{a\to b}^{na}$. After some algebra, in particular using the facts that levels $\varepsilon_{2}$ and $\varepsilon_{3}$ are `idle', and also the mirror symmetry between levels $\varepsilon_{1}$ (resp. $\varepsilon_{2})$ and $\varepsilon_{4}$, (resp. $\varepsilon_{3})$, we obtain 
\begin{equation}
\begin{split}
 \rm{W}_{1\to2}^{ad} &= \rm{f}^{(1)}\left(\varepsilon_4^{(1)}- \varepsilon_4^{(2)}\right),\\
 \rm{W}_{1\to2}^{na} &= 2 (1-\rm{P}) \rm{f}^{(1)} \varepsilon_4^{(2)},
 \end{split}
\end{equation}
\begin{equation}
\begin{split}
 \rm{W}_{2\to1}^{ad} &= \rm{f}^{(2)}\left(\varepsilon_4^{(2)}- \varepsilon_4^{(1)}\right),\\
 \rm{W}_{2\to1}^{na} &= 2 (1-\rm{P}) \rm{f}^{(2)} \varepsilon_4^{(1)},
\end{split}
\end{equation}
where we have defined the `work function'
\begin{equation}\label{eq:f}
 \rm{f}^{(j)} \coloneqq \rm{p}_{1}^{(j)} - \rm{p}_{4}^{(j)} = \frac{ \sinh \beta_{j} \varepsilon_4^{(j)} }{\cosh\beta_{j}\varepsilon_3 + \cosh \beta_{j}\varepsilon_4^{(j)} }.
\end{equation}

Notice that we omitted the upper index $j$ on $\varepsilon_3$ due to the fact that it is independent of $\rm{h(t)}$. The adiabatic contributions to the work exchanges have different signs in each stroke: $\rm{W}_{1\to2}^{ad}\geq0$, but $\rm{W}_{2\to1}^{ad} \leq 0$. Meanwhile, the non-adiabatic contributions are always positive. Thus, as previously mentioned, the nonadiabaticity will tend to increase the work invested in the compression stroke, but decrease the work extracted in the expansion stroke. 

It is interesting to ask whether we can ever have $\rm{W}_{2\to1}^{na} > |\rm{W}_{2\to1}^{ad}|$, i.e., whether the nonadiabaticity may be strong enough to invert the sign of $\rm{W}_{2\to1}$ as a whole. A short calculation shows that such an inversion would require
\begin{equation} \label{Pbound}
\rm{P} < \frac{1}{2}\left[1+ \sqrt{\frac{\varepsilon_4^{(2)}}{\varepsilon_4^{(1)}}}\right] = \frac{1}{2}\left[1+ \sqrt{\frac{4\rm{h_{2}}^2+(\rm{J}_x-\rm{J}_y)^2}{4\rm{h_{1}}^2+(\rm{J}_x-\rm{J}_y)^2}}\right].
\end{equation}
Comparing however with Eq. (\ref{Pinstant}), and remembering that $\rm{h_{1} > h_{2}}$, we see that $\rm{P}_{0}$ is greater than the upper bound in Eq.\,(\ref{Pbound}). Thus we can be sure that work continues to be extracted during very short nonadiabatic expansion strokes. Furthermore if, as suggested by Fig.\,\ref{C_J1_J2},  $\rm{P}(\tau)$ always attains its minimum value in the limit where $\tau \to 0$, then the same would hold for expansion strokes of any duration. We conjecture this is always true for any protocol with $\rm{h(t)}$ monotonically increasing with $t$, and perhaps even for arbitrary protocols, but have not been able to prove or disprove either statement.

The overall work Eq.\,(\ref{Wconvention}) can also be split in the form $\rm{W_{cyc} = W_{cyc}^{ad}+ W_{cyc}^{na}}$ with
\begin{equation}\label{Wcyc}
\begin{split}
 \rm{W}_{cyc}^{ad} & =  \rm{W}_{1\to2}^{ad}+ \rm{W}_{2\to1}^{ad}=\left(\rm{f}^{(1)}-\rm{f}^{(2)}\right)\left(\varepsilon_4^{(1)}- \varepsilon_4^{(2)}\right),\\
 \rm{W}_{cyc}^{na} & = \rm{W}_{1\to2}^{na}+\rm{W}_{2\to1}^{na}= 2 (1-\rm{P}) \left(\rm{f}^{(1)} \varepsilon_4^{(2)} +  \rm{f}^{(2)} \varepsilon_4^{(1)}\right).
\end{split}
\end{equation}

Finally, the heat exchanges $\rm{Q_{1}}$ and $\rm{Q_{2}}$ are
\begin{equation}\label{eq:heats}
\begin{split}
\rm{Q_{1}} & = \mathbb{E}_1 - \mathbb{E}_4 = \mathbb{E}_1 -  \mathbb{E}_3 - \rm{W_{2\to1}},\\
\rm{Q_{2}} & = \mathbb{E}_3 - \mathbb{E}_2  = \mathbb{E}_3 -  \mathbb{E}_1 -\rm{W_{1\to2}}.
\end{split}
\end{equation}

Clearly, the effect of nonadiabaticity on each heat transfer comes exclusively from the nonadiabatic work in the immediately preceding unitary stroke:
\begin{equation}
\begin{split}\label{nonadheat}
\rm{Q_{1}^{na}} & = - \rm{W_{2\to1}^{na}},\\
\rm{Q_{2}^{na}} & = - \rm{W_{1\to2}^{na}}.
\end{split}
\end{equation}
In particular, these contributions are always negative, i.e., correspond to increased dissipation into both baths.

Meanwhile, the adiabatic contributions to each of these quantities are, respectively
\begin{equation}
\begin{split}
\rm{Q_{1}^{ad}} & = \left(\rm{\Delta p}_{3} - \rm{\Delta p}_{2} \right) \varepsilon_{3} +  \left(f^{(2)} - f^{(1)} \right)\varepsilon_{4}^{(1)}, \\
\rm{Q_{2}^{ad}} & = \left(\rm{\Delta p}_{2} - \rm{\Delta p}_{3} \right) \varepsilon_{3} +  \left(f^{(1)} - f^{(2)} \right)\varepsilon_{4}^{(2)}. 
\end{split}
\end{equation}
where $\rm{\Delta p}_{j} \coloneqq p_{j}^{(1)} - p_{j}^{(2)}$. 
Either of these expressions may be positive or negative.

\subsection{\label{regimes}Regimes of Operation}
The laws of thermodynamics allow the Otto cycle to operate in one of four possible thermodynamical regimes  \cite{Buffoni19, PhysRevB.101.054513}, 
depending on the temperatures $\rm{T}_1,\rm{T}_2$, and on the signs of $\rm{Q}_1, \rm{Q}_2$ and $\rm{W}_{\rm{cyc}}$. The first two of these regimes are the usual heat engine, where heat is absorbed from the hot bath and partly converted into work, with the remainder dumped into the cold bath, and the refrigerator, where a source of work is used to remove heat from the cold bath and dump it into the hot one. It is also possible to have an \emph{accelerator}, where a source of work is used to remove heat from the hot bath and release a greater amount of heat into the cold one. Finally, there exists the \emph{heater} regime, where a source of work is used to dump heat into both baths.

  \onecolumngrid
 
 \begin{figure}[t] 
	\includegraphics[scale=.48 ]{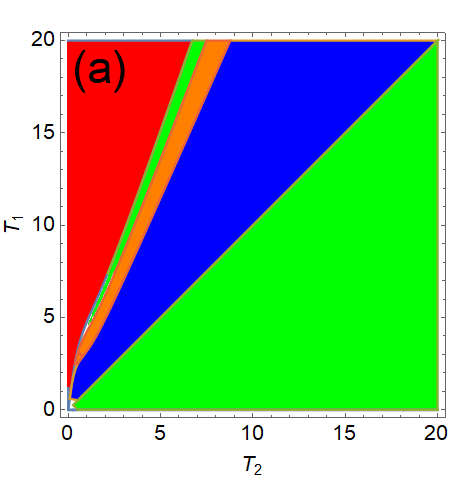}
	\includegraphics[scale=.48 ]{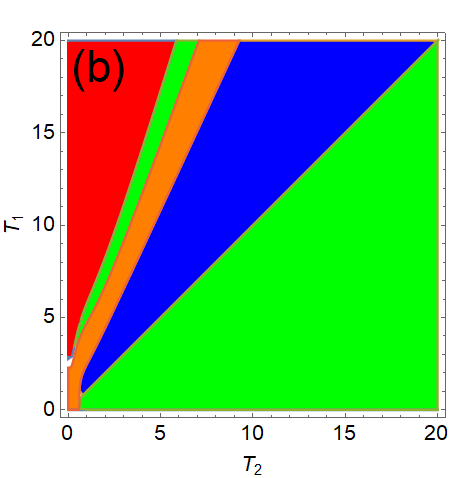}
	\includegraphics[scale=.36 ]{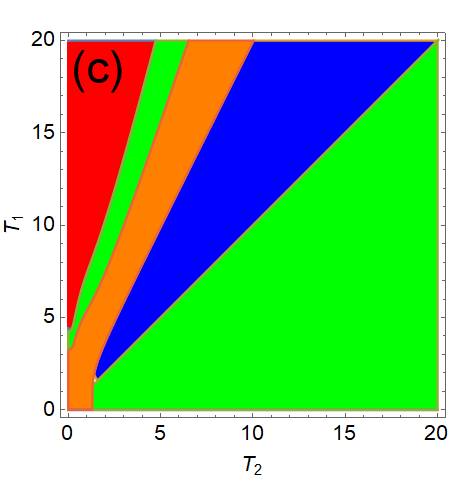}
	\includegraphics[scale=.48 ]{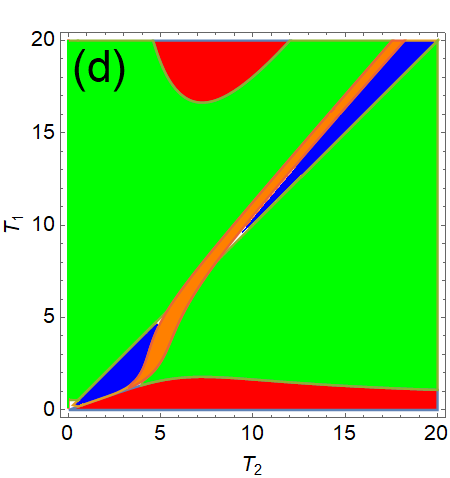}
	\includegraphics[scale=.48 ]{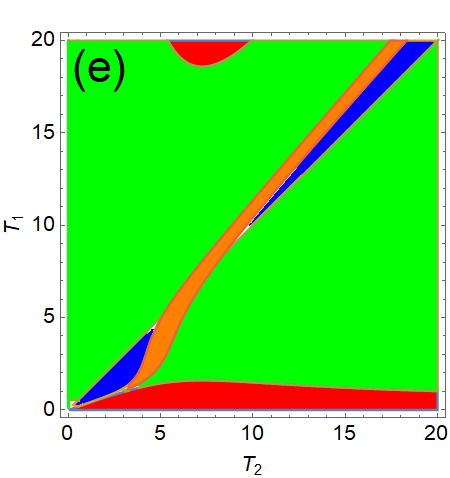}
	\includegraphics[scale=.48 ]{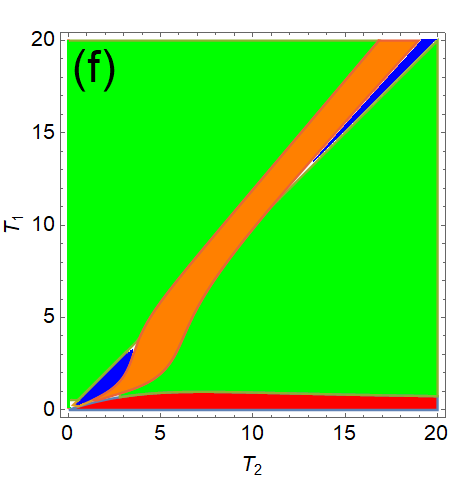}
	\centering
\caption{\label{j1_j2_} Temperature ranges where the cycle operates as a refrigerator (blue/darkest gray),  engine (red/darker gray), heater (orange/lighter gray), or accelerator (green/lightest gray), for different values of the adiabaticity $\rm{P}$. In the top row: $\rm{J}_x=0.01$, $\rm{J}_y=2.00$, $\rm{h}_1=4$ and $\rm{h}_2=1$, with $\rm{P}=1.0$, $\rm{P}=0.97$ and $0.93$ from left to right. In this parameter range only regular engines (where $\rm{T}_1 > \rm{T}_2$) occur. In the bottom row  $\rm{J}_x=10.0$, $\rm{J}_y=2.6$, $\rm{h}_1=4$ and $\rm{h}_2=1$, with $\rm{P}=1.0$, $\rm{P}=0.993$ and $0.95$ from left to right. In this case, Eq.\,(\ref{CRcondition}) is satisfied, leading to the presence of both the regular and counter-rotating engine regimes.}
\end{figure}

\twocolumngrid

As was pointed out in \cite{deoliveira2020efficiency}, for a quantum Otto cycle all of these regimes of operation are in general possible, regardless of the sense of rotation in which the cycle is performed, i.e, regardless of whether $\rm{T_{1} > T_{2}}$ or vice-versa. For each of these cases, the conditions defining each regime may be summarized as follows:
\begin{equation}
    \begin{split}
        &(\rm{T}_1>\rm{T}_2)\\
        \rm{Engine:}&\hspace{.2cm}\rm{Q}_1>0,\hspace{.2cm}\rm{Q}_2<0,\hspace{.2cm}\rm{W}_{\rm{cyc}}<0\\
       \rm{Refrigerator:}&\hspace{.2cm}\rm{Q}_1<0,\hspace{.2cm}\rm{Q}_2>0,\hspace{.2cm}\rm{W}_{\rm{cyc}}>0\\
        \rm{Accelerator:}&\hspace{.2cm}\rm{Q}_1>0,\hspace{.2cm}\rm{Q}_2<0,\hspace{.2cm}\rm{W}_{\rm{cyc}}>0\\
        \rm{Heater:}&\hspace{.2cm}\rm{Q}_1<0,\hspace{.2cm}\rm{Q}_2<0,\hspace{.2cm}\rm{W}_{\rm{cyc}}>0
	    \end{split}
\end{equation}
\begin{equation}
	    \begin{split}
	        &(\rm{T}_1<\rm{T}_2)\\
	        \rm{Engine:}&\hspace{.2cm}\rm{Q}_1<0,\hspace{.2cm}\rm{Q}_2>0,\hspace{.2cm}\rm{W}_{\rm{cyc}}<0\\
	        \rm{Refrigerator:}&\hspace{.2cm}\rm{Q}_1>0,\hspace{.2cm}\rm{Q}_2<0,\hspace{.2cm}\rm{W}_{\rm{cyc}}>0\\
	        \rm{Accelerator:}&\hspace{.2cm}\rm{Q}_1<0,\hspace{.2cm}\rm{Q}_2>0,\hspace{.2cm}\rm{W}_{\rm{cyc}}>0\\
	        \rm{Heater:}&\hspace{.2cm}\rm{Q}_1<0,\hspace{.2cm}\rm{Q}_2<0,\hspace{.2cm}\rm{W}_{\rm{cyc}}>0
	    \end{split}
	\end{equation}

Which regime is realized is highly dependent not only on the temperatures $\rm{T}_1$ and $\rm{T}_2$ of the baths, but on the set of parameters $\rm{h}_1,\rm{h}_2,\rm{J}_x,\rm{J}_y$ of the Hamiltonian, and also on the degree of adiabaticity $\rm{P}$ of the cycle. In Fig.\,\ref{j1_j2_} we illustrate some of this dependence. In the upper row, Fig.\,\ref{j1_j2_}(a-c), we show a weak-coupling situation, defined by the relation 
$\rm{J}_x \rm{J}_y < \rm{h}_1^2$ .
The lower row, Fig.\,\ref{j1_j2_}(d-f), displays a strong-coupling situation, where
\begin{equation}\label{CRcondition}
\rm{J}_x \rm{J}_y > \rm{h}_1^2.
\end{equation}
(See below for a justification of this criterion). In both cases the adiabaticity reduces from left to right, starting from the adiabatic limit ($\rm{P} = 1$). The different operation regimes are represented respectively by the following colors/grayscale tones: Refrigerator: blue/darkest grey; Heat engine: red/dark grey; Heater: orange/light gray; Accelerator: green/lightest gray.

\begin{figure}[h]
		\includegraphics[width=0.8\linewidth]{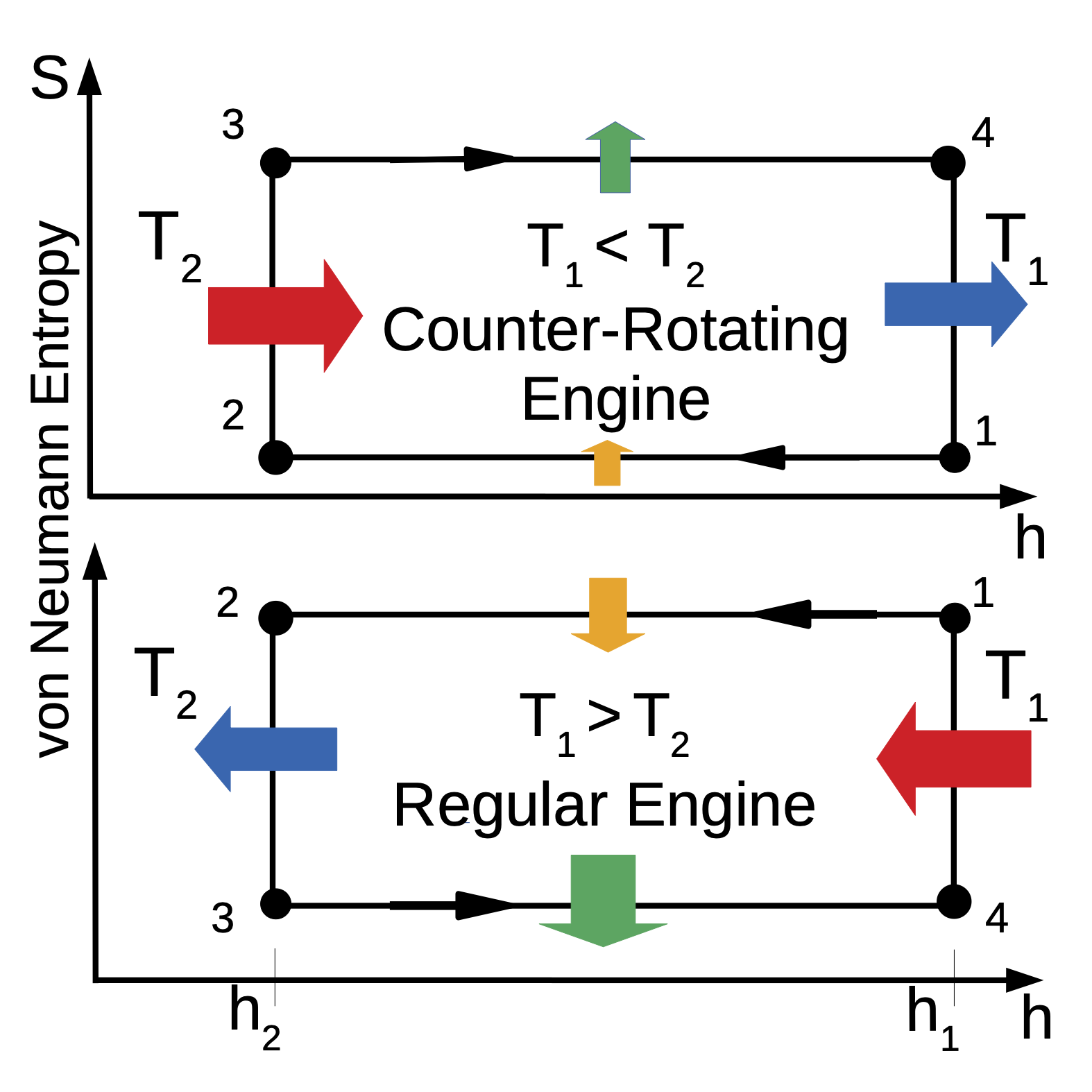}
		\centering
		\caption{\label{fig:s-h-diag} `Regular'  and `Counter-rotating' Otto heat engines, represented as cycles on a von Neumann entropy ($\rm{S}$)  vs field  intensity ($\rm{h}$) diagram (adapted from Ref. \cite{deoliveira2020efficiency}).  }
	\end{figure}

Broadly speaking, in the adiabatic limit (Figs.\,\ref{j1_j2_}(a) and \ref{j1_j2_}(d)), the layout of these diagrams is very similar to the corresponding ones studied in \cite{deoliveira2020efficiency} for the isotropic Heisenberg interaction. Let us now describe and discuss their significant qualitative features. 
Note first that, in the weak-coupling case (Fig.\,\ref{j1_j2_}(a)), there is only one `engine' zone, occupying part of the region where $\rm{T_{1} > T_{2}}$, and including the axis $\rm{T_{2}} = 0$ (i.e., the cycle always operates as an engine when $\rm{T_{2}}\to 0$). We refer to this type of engine cycle as a `regular' engine. It can be represented by an anti-clockwise cycle on an $S \times h$ diagram, where $S$ is the von Neumann entropy  \cite{deoliveira2020efficiency} (Fig. \ref{fig:s-h-diag}, bottom). As $\rm{T_{2}} $ is increased, for fixed $\rm{T_{1}}$, other zones, appear in sequence, corresponding to the accelerator, heater, refrigerator and again accelerator regimes.

In the strong-coupling case (Fig.\,\ref{j1_j2_}(d)), new features appear. Most prominently, there are now two separate engine zones (and also two refrigerator zones). 
The new engine zone occurs for $\rm{T_{2} > T_{1}}$ and includes the axis $\rm{T_{1}} = 0$. It can be represented by a clockwise cycle on an $S \times h$ diagram, and thus we refer to it as a `counter-rotating' engine (Fig. \ref{fig:s-h-diag}, top). In other words, depending on the external bath temperatures, this spin system can function as a heat engine in either sense of rotation on the Otto cycle. 
The two engine zones are separated by a `temperature gap' for $\rm{T_{1}}$, within which no heat engine is possible. Moreover, 
even for $\rm{T_{1}}$ outside this gap, a `regular' engine also ceases to be possible for sufficiently low $\rm{T_{2}} $. In other words, unlike in the weak-coupling case, here the cold bath can always becomes too cold for a `regular' engine to run. Similarly, a  `counter-rotating' engine ceases to be possible for sufficiently high $\rm{T_{2}}$ - i.e., if the hot bath becomes too hot. 

It is not hard to understand physically why  Eq.\,(\ref{CRcondition}) is a sufficient condition for `counter rotating' engines to exist, at least in the adiabatic limit \footnote{In Appendix \ref{Formal-analysis} we show it is in fact also a necessary condition}. Note first that this inequality is equivalent to  $\varepsilon_{2} < \varepsilon_1^{(1)}$. This means that the eigenstate $|\varepsilon_2\rangle$ is in fact the ground state while the system is equilibrating with the cold bath at temperature $\rm{T}_1$. Thus, after equilibration most of the population gets concentrated in this state. But since this is an `idle' level that does not shift nor couple to any others during the subsequent `compression' stroke from $\rm{h}_1\to\rm{h}_2$, it does not make any contribution to the injected work $\rm{W}_{1\to2}^{ad} > 0$.  Thus, for sufficiently low $\rm{T_{1}}$,  $\rm{W}_{1\to2}^{ad}$ must tend to zero. Meanwhile, as we have noted above, the work exchange $\rm{W}_{2\to1}$ during the expansion stroke is always $ <0 $, in the adiabatic limit. Thus, for sufficiently low $\rm{T_{1}}$, we must  have $\rm{W}^{ad}_{cyc} <0$, i.e., a heat engine must exist.

Figs.\,\ref{j1_j2_}(b,c) and \,\ref{j1_j2_}(e,f), illustrate what happens  as the degree of adiabaticity $\rm{P}$ decreases, for weak- and strong-coupling situations respectively. In all figures, $\rm{P} \geq P_{0}$, where the latter is given by Eq.\,(\ref{Pinstant}). (Note again that, even for these `quench strokes', $\rm{P}$ can be quite close to 1). In both situations the engine and refrigerator regions shrink, while the heater zone grows. The reason for this is evident from Eqs.\,(\ref{Wcyc}) and (\ref{nonadheat}): as we have already remarked, the nonadiabatic contributions reduce the work output of the cycle, and also increase dissipation into the baths. As these contributions increase (equivalently, as $\rm{P}$ decreases), they can eventually surpass the adiabatic terms that allowed an engine (respectively, refrigerator) to exist. 

For weak coupling, this effect is strongest at low temperatures: in particular, we can see that, Figs.\,\ref{j1_j2_}(b,c), the `engine zone' effectively shifts upward, i.e., it can only exist above a minimum threshold in $\rm{T_{1}}$. Similarly, for strong coupling the `regular' engine zone also shifts upward in Fig.\,\ref{j1_j2_}(e) (in Fig.\,\ref{j1_j2_}(f), it has shifted entirely out of the depicted range). Meanwhile, the `counter-rotating' engine zone is squashed down, effectively increasing the `temperature gap'. As $\rm{P}$ becomes smaller, we expect these shifts to become more and more pronounced. In the limit where $\rm{P}<0.5$,  Eq. (\ref{Wcyc}) implies that  $\rm{W_{cyc}} > 0$, i.e, an engine ceases to be possible at all.
Nevertheless, the important thing to note is that essentially all the features of the adiabatic limit continue to be present when $0.5< \rm{P}<1$. In other words, nonadiabaticity does not significantly alter the qualitative results reported in \cite{deoliveira2020efficiency}. This is one of the main conclusions of this paper.
 
It is worth remarking that, as was pointed out in \cite{deoliveira2020efficiency}, an Otto cycle cannot behave as a heater unless the working substance leaves thermal equilibrium during the unitary strokes. If it remains in a thermal state, it necessarily loses entropy while ceding heat to a heat bath. Thus, it cannot cede heat to both baths, as then it would not be able to return to its initial state at the end of the cycle. This is why, for example, a single-qubit Otto cycle such as the one studied in \cite{PhysRevB.101.054513} cannot behave as a heater in the adiabatic limit: in that limit, the qubit remains in an energy-diagonal (and hence thermal) state during the unitary strokes. For this system, an escape from equilibrium can only arise due to nonadiabaticity, which introduces energy-basis coherence. This allows a heater to occur \cite{PhysRevB.101.054513}. In contrast, for systems of larger dimension, such as the one we study here, a heater may occur even in the adiabatic limit. This is because thermal states with respect to $\hat{\rm{H}}_1$ do not in general evolve adiabatically into thermal states with respect to $\hat{\rm{H}}_2$. (This only happens if all level gaps change proportionately during the adiabatic stroke - something which, for example, is generally not true in the presence of idle levels \cite{deoliveira2020efficiency}). Moving away from the adiabatic limit allows even greater deviations from equilibrium, leading to an increase in dissipation into the baths during the equilibration strokes. As a result, we can expect the heater regime to becomes more and more prevalent as the degree of adiabaticity is decreased. Indeed, this is observed both in \cite{PhysRevB.101.054513}, and also here. Nevertheless, this is not always necessarily the case: within carefully chosen cycle parameter ranges, the other thermodynamical regimes, including regular and counter-rotating engines, remain possible even for  $\rm{P}$ close to $0.5$ (see section \ref{sec:deepAR} below).

So far we have only discussed Fig.\,\ref{j1_j2_}(b,c) in qualitative terms. A quantitative analysis is also possible to a certain extent. In Appendix \ref{Formal-analysis} we formally prove three analytical results relating to the features we have described above.  Result 1 discusses the conditions for `regular' heat engines. First of all we show that, for weak coupling, such engines are always possible  when $\rm{T_{2}} \to 0 $ in the adiabatic limit. We also prove that, for cycles with non-adiabatic strokes, this remains true but only for $\rm{T_{1}}$ above a certain threshold. In contrast, for strong coupling, a heat engine ceases to be possible as $\rm{T_{2}}\to 0$.  Result 2 is concerned with the conditions for `counter-rotating' engines. We show that the strong-coupling condition in Eq.\,(\ref{CRcondition}) is indeed necessary and sufficient for the counter-rotating engine regime to be possible, both in the adiabatic limit and in the case of very short nonadiabatic evolutions - and in fact for any nonadiabatic cycle with $\rm{P} \geq \rm{P}_{0}$. Finally, Result 3 proves the 
existence of the temperature gap for $\rm{T_{1}}$ in the strong-coupling situation. In other words, there  always exists a range of values $\rm{T_{1}} \in (\rm{T_{1}^{a}, T_{1}^{b}})$, inside which the cycle cannot operate at all as a heat engine - whatever the value of $\rm{T_{2}}$. For $\rm{T_{1}}$ within this range, the work extracted during the expansion stroke of the cycle is always less than the one invested in the compression stroke. Furthermore, this gap is in fact `direct', i.e., the peak of the `counter-rotating' engine zone is located at the same value of $\rm{T_{2}}$ as the lowest point of the `regular' zone. 

With regard to the boundaries between the different zones, it should be clear that, even in the adiabatic case, no simple analytical expressions can describe them entirely, except in particular limits. For example, for completely uncoupled spins ($\rm{J}_x = \rm{J}_y = 0)$, it can be shown
that: for $\rm{T}_1 > (\rm{h}_1/\rm{h}_2) \rm{T}_2$ we have an engine; for $(\rm{h}_1/\rm{h}_2) \rm{T}_2 > \rm{T}_1 > \rm{T}_2$ we have an refrigerator and for $\rm{T}_1 < \rm{T}_2$ we have an accelerator.  Fig. \ref{j1_j2_}(a), where the coupling is small but nonzero, approximates this limit.  Nevertheless, for any amount of coupling, it is possible in principle to obtain asymptotic expressions for the various zone boundaries, valid for large enough or small enough $\rm{T_{1}}$ and $\rm{T_{2}}$. We do not give these details here, however. See \cite{deoliveira2020efficiency} for a similar calculation.

\subsubsection{Counter-rotating engines in the deep nonadiabatic regime} \label{sec:deepAR}

In the nonadiabatic examples illustrated in Fig. \ref{j1_j2_}, the adiabaticity parameter $\rm{P}$ remained always above 0.9, a value still not too far from the adiabatic limit. It is reasonable to wonder then whether features such as counter-rotating engines (CREs) can still exist in the `deep nonadiabatic regime', where $\rm{P}$ approaches 0.5 \footnote{Recall that, as seen in the last section, no engine can exist if $\rm{P}<0.5$}. In this section, we show that this is indeed the case. 

First though, let us briefly discuss what it takes to attain this regime. It is important to realize that $\rm{P}$ is not an independent variable, but is constrained by the cycle parameters $\rm{h}_1,\rm{h}_2,\rm{J}_x,\rm{J}_y$, and by the choice of protocol $\rm{h(t)}$. In our examples, these choices were such that $\rm{P}$ turns out to remain above 0.9 for all stroke durations, even very short ones (similarly to Fig. \ref{C_J1_J2}). This illustrates the important point that the limit of small $\rm{P}$ is not the same as the one defined by very fast strokes.

It is conceivable that, for some different protocol $\rm{h(t)}$, $\rm{P(\tau)}$ could perhaps become small. However, note that, for any choice of $\rm{h(t)}$, $\rm{P}$ must have the same limiting value $\rm{P_{0}} \equiv \rm{P(\tau \to 0)}$ for very short strokes, given by Eq. (\ref{Pinstant}), and must reach 1 for very long strokes, as discussed in Section \ref{non_adiabatic}.

Thus, for the cycle parameters used in Fig. \ref{j1_j2_}, $\rm{P(\tau)}$ could only conceivably become small for some particular choice of protocol and some finite range of stroke durations $\tau$. In fact though, we could not construct any example of a protocol for which $\rm{P}(\tau)$ ever dipped below $\rm{P_{0}}$.  We conjecture (but have not been able to prove either way) that no such protocol exists.

One can of course always just input `by hand' a small value of $\rm{P}$  into the equations for work and heat, and analyze the consequences  \cite{PhysRevB.101.054513}. However, it is by no means clear whether doing this is physically consistent, i.e, whether such a $\rm{P}$ is actually achievable without also changing the cycle parameters $\rm{h}_1,\rm{h}_2,\rm{J}_x,\rm{J}_y$. In other words, these consequences may well not be physical.

This is not to say that a small $\rm{P}$  is impossible. In fact, it is not hard to choose values of these parameters for which $\rm{P_{0}}$ itself becomes small. It can be seen from Eq. (\ref{Pinstant}) that a sufficient condition for this is as follows:
\begin{align}\label{DARconds}
2\rm{h}_{1} \gg |\rm{J}_x-\rm{J}_y | \gg  2\rm{h}_{2} ,
\end{align}
in which case we have approximately 
\begin{align}
\rm{P_{0}} - 1/2 \sim \frac{\rm{h_{2}}}{\left|\rm{J}_x-\rm{J}_y\right|} + \frac{\left|\rm{J}_x-\rm{J}_y\right|} {4\rm{h_{1}}} \ll 1
\end{align}

As we have discussed above, a small $\rm{P}$ usually favors the heater regime, since it suppresses extracted work and input heat and enhances both input work and output heat.  Nevertheless, it turns out that a CRE can indeed still occur in this limit.

To see this, we call attention again to Result~2 in Appendix \ref{Formal-analysis}. As we have mentioned, this result shows that, for any sufficiently short nonadiabatic stroke, a CRE will remain possible if, and only if, the strong-coupling condition in Eq.\,(\ref{CRcondition}) is met. Thus, in order to obtain a CRE in the deep adiabatic regime, it is sufficient to find a combination of cycle parameters for which, simultaneously, $\rm{P_{0}}$ is small and Eq.\,(\ref{CRcondition}) is satisfied. One way to achieve this, for instance, is to take a set of values that satisfy the conditions set out in Eq. (\ref{DARconds}), and then increase both $\rm{J}_x$ and $\rm{J}_y$ by equal amounts (preserving the distance $\rm{J_x - J_y}$ ), until Eq.\,(\ref{CRcondition}) is achieved. Note that it is also necessary of course to choose the appropriate bath temperatures. According again to Result 2, a CRE will always occur for low enough $\rm{T_{1}}$ and any $\rm{T_{2}} > \rm{T_{1}}$ ).

\begin{figure}[t] 
	\includegraphics[scale=.48 ]{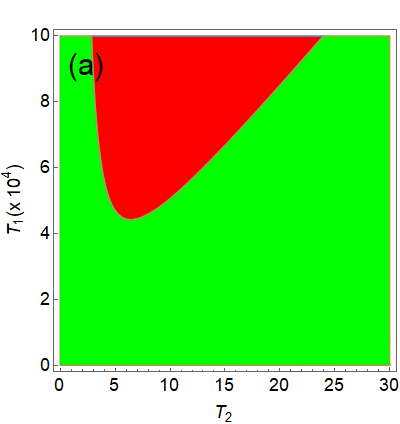}
	\includegraphics[scale=.46 ]{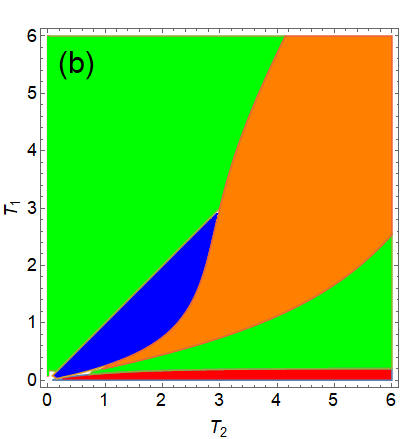}
         \centering
\caption{\label{smallPEngines} Example of a scenario in the `deep adiabatic' regime ($\rm{P} \sim 0.5$), for which nevertheless the Otto cycle may still  operate either as (a) a regular engine or (b) a counter-rotating engine. As in Fig. \ref{j1_j2_}, blue/darkest gray corresponds to the refrigerator regime, red/darker gray areas to the engine, orange/lighter gray to the heater, and green/lightest gray to the accelerator. In both figures,  $\rm{J}_x=5.0$, $\rm{J}_y=4.9$, $\rm{h}_1=4.0$ and $\rm{h}_2=0.001$, while $\rm{P}= \rm{P}(\tau \to 0) = 0.516$. Note the extreme difference between the ranges of $\rm{T_{1}}$ at which each engine regime occurs.}
\end{figure}

A concrete example, for which $\rm{P} = 0.516$, is illustrated in Fig. \ref{smallPEngines}.  As predicted in the previous section, the small value of $\rm{P}$ results in each engine regime occurring in ranges of $\rm{T_{1}}$ that are extremely shifted with respect to one another (note the difference in the scale of $\rm{T_{1}}$ in both graphs).

\section{\label{heat_engine}Efficiency}

The efficiency $\eta = -\rm{W}/\rm{Q_{hot}}$ of a quantum Otto cycle depends on the level structure of its working substance, and on how these levels shift during the unitary strokes. If all energy gaps shift proportionately to each other, and the stroke is performed adiabatically, then $\eta$ equals the `standard Otto efficiency' $\eta_{Otto} = 1-\rm{h}_2/\rm{h}_1$ \cite{PhysRevE.76.031105}. This is the case, for example, for a working substance consisting of uncoupled single qubits \cite{doi:10.1063/1.461951, PhysRevLett.93.140403} or a harmonic oscillator \cite{e19040136}, or even some coupled spin models \cite{PhysRevE.65.055102}. In all these cases, the system remains in a thermal state throughout the entire cycle, and so the quantum adiabatic stroke is also adiabatic in the usual thermodynamical sense  \cite{PhysRevE.76.031105}. 

However, Otto cycles with $\eta > \eta_{Otto}$ are also possible if the working substance is a multi-level quantum system with an appropriate level structure.  As discussed in \cite{deoliveira2020efficiency}, one example of such a structure occurs when a subset of the energy levels are `working' levels that shift proportionately to an external classical driving field, while the remainder are `idle' levels, which do not shift, and therefore do not contribute to work. Such a state of affairs may be induced by an appropriate coupling between subsystems \cite{PhysRevE.83.031135, deoliveira2020efficiency}, but can also occur in a working substance with no subsystems, such as a qutrit \cite{PhysRevE.104.054128}.  In these cases, the efficiency gain relative to $\eta_{Otto}$ can be interpreted as  the result of a reversed partial heat flux via the `idle' levels, flowing from the cold to the hot bath, accompanied by an increase in the ordinary (from hot to cold) heat flux via the `working' levels.  Other related effects also occur, such as the possibility of an increase in $\eta$ when the temperature difference between the baths decreases.

\begin{figure}[t]
		\includegraphics[scale=.67]{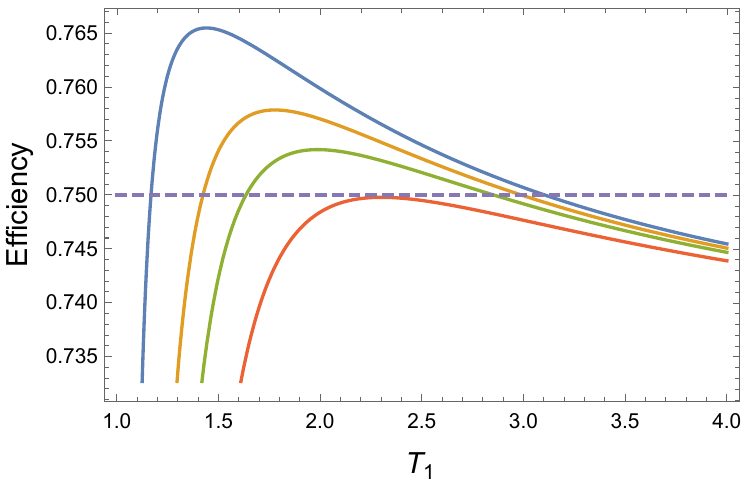}
		\centering
\caption{\label{regular_efficiency_temp_P_small_J_with_inset} Efficiency of a regular engine as a function of the hot bath temperature $\rm{T}_1$, for different values of the adiabaticity parameter $\rm{P}$, with fixed cold bath temperature $\rm{T}_2=0.2$. Cycle parameters are $\rm{h}_1=4.0$, $\rm{h}_2=1.0$, $\rm{J}_x=0.01$ and $\rm{J}_y=0.8$.  From top to bottom: $\rm{P}=1.0$, $\rm{P}=0.9999$, $\rm{P}=0.9998$ and $\rm{P}=0.9996$.  As $\rm{P}$ decreases, the efficiency gets monotonically worse in all cases. Nevertheless, for sufficiently large but not ideal adiabaticity, it can attain values above the standard Otto efficiency $\eta_{Otto}=1-\rm{h}_2/\rm{h}_1=0.75$ (dashed horizontal line), for certain temperature ranges.}  	
\end{figure}

\begin{figure}[t]
	\includegraphics[scale=.675]{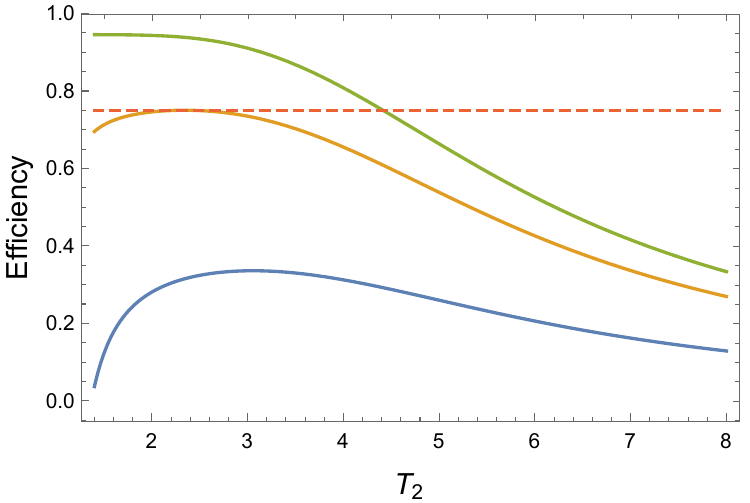}
		\centering
		\caption{\label{efficiency_temp_P_with_inset}Efficiency of a counter-rotating engine as a function of the hot bath temperature $\rm{T}_2$, for different values of the adiabaticity parameter $\rm{P}$, with fixed cold bath temperature $\rm{T}_1=0.05$.   Cycle parameters are: $\rm{h}_1=4.0$, $\rm{h}_2=1.0$, $\rm{J}_x=10.0$ and $\rm{J}_y=1.6$.   Note these satisfy Eq.\,(\ref{CRcondition}), which is the condition for counter-rotating engines to exist for small values of $\rm{T}_1$.  From top to bottom. $\rm{P}=1.0$, $\rm{P}=0.9785$, and $\rm{P = P(\tau\to 0)} = 0.932$.  As $\rm{P}$ decreases, the efficiency gets monotonically worse in all cases. As in Fig. \ref{regular_efficiency_temp_P_small_J_with_inset}, for sufficiently large $\rm{P}$, $\eta$ can attain values above the standard Otto efficiency $\eta_{Otto}=1-\rm{h}_2/\rm{h}_1=0.75$ (dashed horizontal line)}

\end{figure}

In the case of the $\rm{XY}$ coupling model discussed in present study, the level structure is similar but not quite equivalent: once again there are `idle' levels ($2$ and $3$) and `working' levels  ($1$ and $4$), but now the latter do not shift proportionately to the external field. Nevertheless, once again $\eta$ may exceed $\eta_{Otto}$ in the adiabatic limit, and once again increases in $\eta$ may occur when the temperature difference between the baths decreases (see Figs. \ref{regular_efficiency_temp_P_small_J_with_inset} and \ref{efficiency_temp_P_with_inset}).

Our chief concern here is to investigate how these effects are impacted by a lack of perfect adiabaticity. As we have previously mentioned, a non-adiabatic evolution draws the system away from thermal equilibrium, since it introduces energy-basis coherence. As a result, we can expect an increase in the energy dissipated as the system re-equilibrates with the heat baths, resulting in a decrease in engine efficiency. Indeed, it is known that, for an Otto engine based on a simple 1-qubit working substance, the efficiency increases monotonically with the degree of adiabaticity $\rm{P}$ \cite{PhysRevB.101.054513}. Another study \cite{bla} has linked a loss of efficiency in a two-qubit $\rm{XY}$ Otto engine such as ours with the speed of operation of the nonadiabatic unitary strokes  - and also shown how these losses can be partially reversed by means of a `shortcut to adiabaticity' technique.

Here we start by generalizing a result of \cite{PhysRevB.101.054513}, proving that the engine efficiency is indeed a monotonically increasing function of $\rm{P}$ (equivalently, $\eta$ falls as the stroke become less and less adiabatic). This is not entirely trivial,  since Eqs.\,(\ref{Wcyc}) and (\ref{eq:heats}) imply that both the output work and the input heat are monotonically increasing with $\rm{P}$. Thus, at least at first sight, their ratio could in principle decrease or increase with $\rm{P}$. In fact, only the latter case is actually possible. To see this, we first require the following easily checked mathematical fact: Suppose $0~< ~f(x)~\equiv~a(x)/b(x) < 1$, where $a(x), b(x) > 0$ and $a'(x) \geq b'(x) > 0$.  Then $f'(x) > 0$.

Consider now a parameter regime where the cycle functions as a `regular' heat engine in the adiabatic limit, i.e., $\rm{0 < -W^{ad}_{cyc}}0 < \rm{Q^{ad}_{1}} $. In this case the efficiency is 
\begin{align}
 \eta &= \frac{\rm{-W_{cyc}}}{\rm{Q_{1}}} = \frac{\rm{-W^{ad}_{cyc} -W^{na}_{1\to2}(P) -W^{na}_{2\to1}(P) }}{\rm{Q^{ad}_{1} - W^{na}_{2\to1}(P)}}
 \end{align}
Since $\rm{W^{na}_{1\to2}(P)}$ and $\rm{W^{na}_{2\to1}(P)}$ are both positive, monotonically decreasing functions of $\rm{P}$, it is clear that $\eta(\rm{P})$ satisfies the conditions outlined above, and therefore $\eta'(\rm{P}) > 0$. The same argument is also clearly applicable to `counter-rotating' engines, just replacing $\rm{Q_{1}}$ with  $\rm{Q_{2}}$. In fact, it should hold for any Otto engine whose adiabaticity is measurable by a single parameter  $\rm{P}$.

The behavior of $\eta$ as $\rm{P}$ varies is illustrated in Figs. \ref{regular_efficiency_temp_P_small_J_with_inset} and \ref{efficiency_temp_P_with_inset}. In both cases we plot the efficiency as a function of the hot bath temperature, for different values of $\rm{P}$. In Fig.~\ref{regular_efficiency_temp_P_small_J_with_inset} we consider a situation with $\rm{T}_1>\rm{T}_2 $  (`regular' engine) and weak coupling ($\rm{J}_x=0.01$, $\rm{J}_y=0.8$, $\rm{h_{1} = 4, h_{2} = 1}$). As expected, $\eta$ decreases monotonically as the degree of adiabaticity falls. Furthermore, for these parameters and in the adiabatic limit, the cycle can attain an efficiency higher than $\eta_{Otto}$  within a certain temperature range. However the efficiency loss due to nonadiabaticity means that this range rapidly shrinks as $\rm{P}$ falls from 1, vanishing already for a very small deviation, namely for $\rm{P}\leq 0.9996$. Nevertheless, an interesting thing to note is that,  for all values of $\rm{P}$,  $\eta$ does not increase monotonically with the hot bath temperature $\rm{T_{1}}$. Instead, it reaches a maximum and then decreases, even though the temperature difference between the baths is increasing. This is very different from what happens in more familiar cycles, such as Carnot cycles, or Otto cycles in simple uncoupled systems, such as a single qubit, or even a classical ideal gas. As we can see, this unusual behavior survives the presence of nonadiabaticity.

In Fig. \ref{efficiency_temp_P_with_inset} we plot a similar graph for a strong-coupling situation with $\rm{T}_1 < \rm{T}_2$ (counter-rotating engine). The behavior is quite similar to that of the regular engine in weakly-coupled regime. Here  $\rm{J}_x=10.0$, $\rm{J}_y=1.6$, $\rm{h_{1} = 4.0, h_{2} = 1.0}$), a choice that again ensures $\eta > \eta_{Otto}$ in the adiabatic limit, within a finite range of $\rm{T}_2$.  Note that, again, $\eta$ decreases for sufficiently large $\rm{T}_2$, i.e.  as the bath temperature difference increases. Also, once again $\eta$ falls monotonically with $\rm{P}$, although here the condition $\eta > \eta_{Otto}$ survives a slightly wider range ($\rm{P} \gtrsim 0.976$). The lower curve corresponds to the efficiency when $\rm{P = P(\tau\to 0)} = 0.932$, calculated using Eq.\,(\ref{Pinstant}). We can thus see that, as predicted in Result 2 in Appendix \ref{Formal-analysis}, even for instantaneous, quench-like unitary strokes the cycle can still function as a counter-rotating engine for low enough $\rm{T_{1}}$, albeit very inefficiently.

\begin{figure}[t]
		\includegraphics[scale=.33]{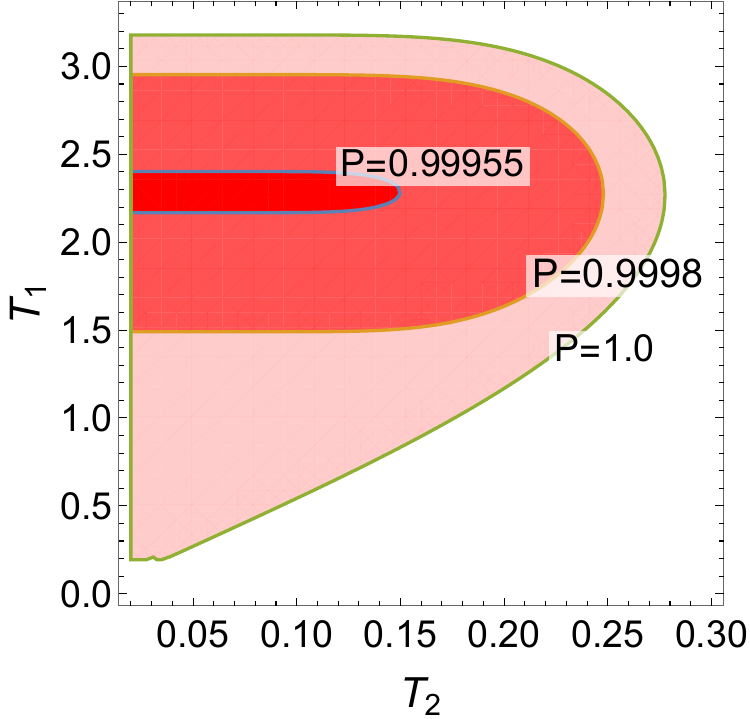}
		\includegraphics[scale=.33]{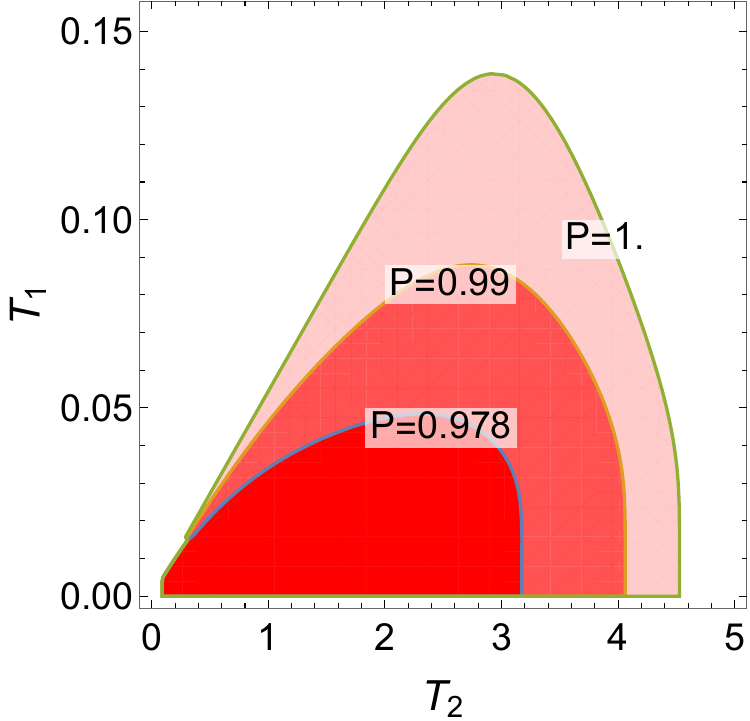}
		\centering
		\caption{\label{regions_with_higher_efficiencies}  Temperature ranges where the cycle operates as a heat engine with efficiency higher than the standard Otto value  $\eta_{Otto} = 1-\rm{h}_2/\rm{h}_1$, for different values of the quantum adiabaticity $\rm{P}$. Left side: a `regular' engine $(\rm{T}_1>\rm{T}_2)$, with $\rm{J}_x=0.01$,  $\rm{J}_y=0.8$. Right side: `counter rotating' engine $(\rm{T}_1<\rm{T}_2)$, with $\rm{J}_x=10.0$,  $\rm{J}_y=1.6$. In both cases, $\rm{h}_1=4.0$ and $\rm{h}_2=1.0$. Note the rapid decrease of these regions as $\rm{P}$ decreases, especially in the `regular' case. Note also the scale of these graphs in comparison with those in Fig. \ref{j1_j2_}.}
\end{figure}

Finally, in Fig. \ref{regions_with_higher_efficiencies} we give a wider view of the temperature ranges within which the engines in Figs. \ref{regular_efficiency_temp_P_small_J_with_inset} and \ref{efficiency_temp_P_with_inset} operate with an efficiency higher than $\eta_{Otto}$. In both cases, we can see how a decrease in  $\rm{P}$ quickly decreases the size of the region with higher efficiency, especially in the case of `regular' engines.

\section{Conclusion and discussion}

One of the main goals of quantum thermodynamics is to understand how the  properties of quantum systems can lead to possible advantages or unusual operational regimes in processes involving energy exchanges. In this work we have investigated quantum Otto cycles containing finite-time unitary strokes, generated by a multilevel Hamiltonian with a time-dependent eigenbasis, resulting in a loss of quantum adiabaticity. We have studied if interesting features described in Ref. \cite{deoliveira2020efficiency} in an adiabatic context are still present in these nonadiabatic cycles, and the extent to which they can endure. We found that qualitative features, such as counter-rotating heat engine cycles, or an engine efficiency that can increase as the baths' temperature difference decreases, are surprisingly resilient to a loss of of quantum adiabaticity, and are still present even for very short, quench-like nonadiabatic strokes.

On the other hand, we also found that quantitative features, such as the possibility of reaching heat engine efficiency beyond the standard Otto value $\eta_{Otto}$, can be much more sensitive. More generally, as the degree of quantum adiabaticity is reduced, we find a general decrease in efficiency in the engine regime. As a consequence, the parameter regions where the cycle operates as an engine contract. In particular, the region where $\eta > \eta_{Otto}$ is rapidly degraded, vanishing even for very a small degree of nonadiabaticity. Similar effects occur for the refrigerator and accelerator regimes. Meanwhile, the heater regime increases in scope, reflecting the greater dissipation needed to rethermalize the system after the non-adiabatic strokes take it further out of equilibrium in comparison with quantum adiabatic ones.

It would be interesting to further investigate if and how these conclusions may be altered when other imperfections and features relevant for a real-world implementation of such a cycle are taken into account. For example, the possible effects of incomplete, finite-time thermalization strokes, or the scenarios in which it may be appropriate to go beyond average work and heat, and consider the large fluctuations in these energy exchanges. Such considerations would allow a more complete and realistic assessment of the role of quantum properties in the power vs efficiency trade-off in these cycles.

\section{Acknowledgements}
This work is supported by the Instituto Nacional de Ciência e Tecnologia de Informação Quântica (465469/2014-0), and by the Air Force Office of Scientific Research under award number FA9550-19-1-0361.

\appendix

\section{Solution of H(t)} \label{Apdx-A}

The Hamiltonian in Eq. (\ref{ws_hamiltonian}), which we reproduce here for the reader's convenience, is
\begin{equation}
\hat{\rm{H}}(\rm{t})=\rm{J}_x\hat{\sigma}_{1x}\hat{\sigma}_{2x}+\rm{J}_y\hat{\sigma}_{1y}\hat{\sigma}_{2y}+h(t)[\hat{\sigma}_{1z}+\hat{\sigma}_{2z}],
\end{equation}
Note first  that $\hat{\rm{H}}(\rm{t})$ conserves both the total spin $\hat{\sigma}^{2}$ and the product $\hat{\sigma}_{1z}\hat{\sigma}_{2z}$. The first of these symmetries implies that, regardless of the values of $\rm{h(t)}, \rm{J}_x$ or $\rm{J}_y$, the singlet state $\left(|\uparrow \downarrow \rangle - |\downarrow \uparrow \rangle\right)/\sqrt{2} \equiv |\varepsilon_2\rangle$ must be an eigenstate of $\hat{\rm{H}}(\rm{t})$. The second symmetry then requires the same to be true for $\left(|\uparrow \downarrow \rangle + |\downarrow \uparrow \rangle\right)/\sqrt{2} \equiv |\varepsilon_3\rangle$. The corresponding eigenvalues are $\varepsilon_3= (\rm{J}_x+\rm{J}_y)$, $\varepsilon_2=-\varepsilon_3$. The problem thus reduces to diagonalizing the reduction of $\hat{\rm{H}}(\rm{t})$ to the 2-dimensional subspace generated by the states $|\uparrow \uparrow \rangle$ and $|\downarrow \downarrow \rangle$. The result of this straightforward exercise is given in Eqs.(\ref{epsilon1}) and (\ref{alphapm}).

\section{Formal Analysis} \label{Formal-analysis}

In this Appendix we analyze more closely the properties of the  `work function' $\rm{f^{(j)}}$ defined in Eq.\,(\ref{eq:f}), and use them to arrive at general conclusions concerning the behavior of the thermodynamic cycle. 

We begin by noting that  $\rm{f^{(j)}}$ may be expressed in a simpler form as 

\begin{equation}\label{eq:fg}
\rm{f^{(j)}}= g\left(\beta_{j}\varepsilon_{4}^{(j)},  \varepsilon_{3}/\varepsilon_{4}^{(j)} \right),
\end{equation}
where
\begin{align}
g(x,y) = \frac{e^{x} - e^{-x}}{e^{xy} + e^{-xy} + e^{x} + e^{-x}}. \label{eq:g}
\end{align}
As was the case for a slightly different function studied in the Appendix to Ref. \cite{deoliveira2020efficiency}, this function has the following easily checked properties  
for $x,y \geq 0:$
\begin{align}
0  \leq g(x,y) &< 1 \label{eq:gpositive}; \;\;\;g(x,y) =0 \iff x = 0  \\
\lim_{x\rightarrow \infty} g(x,y) & = 
\begin{cases}
1, & 0\leq y < 1; \\
1/2, & y = 1; \\
0, & y > 1. \\
\end{cases}\label{eq:limitg}\\
\frac{\partial g(x,y)}{\partial x} &> 0 \text{ for $0 \leq y \leq 1$.} \label{eq:monotx}\\
\frac{\partial g(x,y)}{\partial x} &= 0 \text{ at a single value of $x$, for $y > 1$} \label{eq:maximumg}\\
\frac{\partial g(x,y)}{\partial y} & < 0.  \label{eq:monoty} \\
g(rx, y) &> g(x,ry) \text{ for $r > 1$.} \label{eq:gscaling}
\end{align}

\medskip

Let us now apply these properties to understanding some relevant features of the thermodynamic cycle.\smallskip

\textbf{Result 1: [Conditions for `regular' heat engines]} (i) Let $\rm{T_{1}}$ be fixed. For $\varepsilon_{3}/\varepsilon_{4}^{(1)} \leq 1$ (weak coupling), we always have $\rm{W^{ad}_{cyc}}<0$ for sufficiently low $\rm{T_{2} < T_{1}}$. In other words, if the cold bath is sufficiently cold, a `regular' heat engine is always possible in the adiabatic limit. (ii) For $\varepsilon_{3}/\varepsilon_{4}^{(1)} > 1$ (strong coupling, equivalent to Eq.\,(\ref{CRcondition})), we always have $\rm{W^{ad}_{cyc}} > 0$ for sufficiently low $\rm{T_{2} < T_{1}}$. In other words, if the cold bath is sufficiently cold, a `regular' heat engine ceases to be possible in the adiabatic limit.  (iii)  For cycles with non-adiabatic strokes, (ii) is still true for the full work $\rm{W_{cyc}}$, but (i) only holds for $\rm{T_{1}}$ above a certain threshold value $\rm{T_{1}^{0}}$.  For $\rm{T_{1} 
\leq T_{1}^{0}}$, no heat engine is possible even for $\rm{T_{2}}\to 0$.  Finally, if $\rm{P} \to \frac{1}{2}\left(1+ \varepsilon_{4}^{(2)}/\varepsilon_{4}^{(1)}\right) \coloneqq P_{\min}$, then $\rm{T_{1}^{0}} \to \infty $. In other words, for $\rm{P} \leq \rm{P_{\min}} $ no engine is possible at all. \medskip

\emph{Remark:} Note $\rm{P_{\min}} $ is actually smaller than $\rm{P}(\tau \to 0)$ in Eq.\,(\ref{Pinstant}). As we have already mentioned, we conjecture that  in fact $\rm{P(\tau) \geq P(\tau\to 0)}$ for any protocol $\rm{h(t)}$. If this is correct, then there will in fact always be a finite threshold value $\rm{T_{1}^{0}}$. \medskip

\emph{Proof}: 
For (i), note that, by Eqs.\,(\ref{eq:fg}) and (\ref{eq:limitg}), $\lim_{\beta_{2}\rightarrow \infty }\rm{f^{(2)}} = 1$, while by Eq.\,(\ref{eq:gpositive}), $0 < \rm{f^{(1)}} < 1$. Thus, for sufficiently low $\rm{T_{2}}$ we must have $\rm{f^{(1)}} - \rm{f^{(2)}} <0$. By Eq.\,(\ref{Wcyc}), this means $\rm{W_{cyc}^{ad}} < 0$.  On the other hand, in the strong-coupling case  Eq.\,(\ref{eq:limitg}) implies that we now have $\lim_{\beta_{2}\rightarrow \infty }\rm{f^{(2)}} = 0$. Thus $\rm{W_{cyc}^{ad}} > 0$ for sufficiently low $\rm{T_{2}}$, and we have (ii). In fact, since $\rm{W_{cyc}^{na}}$ is also $>0$, we can immediately conclude that $\rm{W_{cyc}} >0$ also in the nonadiabatic case. To finish proving (iii), we need to verify that (i) continues to hold even for nonadiabatic cycles, but only for sufficiently high $\rm{T_{1}}$. Let us begin by noting that, when the nonadiabatic term in Eq.\,(\ref{Wcyc}) is taken into account, the sign of $\rm{W_{cyc}}$ is no longer determined simply by whether $\rm{f^{(1)} < f^{(2)}}$. Instead, a short calculation shows that the necessary and sufficient condition for $\rm{W_{cyc} }<0$ is now 
\begin{equation}\label{nonadiabatic-cond}
\rm{f^{(1)} < c(P) f^{(2)}},
\end{equation} 
where
\begin{equation}
 \rm{c(P)} \coloneqq \frac{\varepsilon_{4}^{(1)}-\varepsilon_{4}^{(2)}-2(1-P)\varepsilon_{4}^{(1)}}{\varepsilon_{4}^{(1)}-\varepsilon_{4}^{(2)} + 2(1-P)\varepsilon_{4}^{(2)}}
\end{equation}
Note $\rm{c(P)}$ is a monotonically increasing function of $\rm{P}$, that tends to 1 as $\rm{P}\to 1$, and tends to $0$ as $\rm{P \to P_{\min}}$. Since $\rm{f^{(j)}>0}$, Eq.\,(\ref{nonadiabatic-cond}) cannot be satisfied for $\rm{P} \leq \rm{P_{\min}} $.  In other words, no engine is possible in this case.

Assuming then $\rm{c(P)}>0$, and using again the fact that, for weak coupling, $\lim_{\beta_{2}\rightarrow \infty }\rm{f^{(2)}} = 1$, condition (\ref{nonadiabatic-cond}) becomes $\rm{f^{(1)}} < c(P)$ in the limit of small $\rm{T_{2}}$. This upper bound is $<1$ when $\rm{P} <1$. Since however   $\rm{f^{(1)}}\to 1$ as $\rm{T_{1}}$ also becomes small, the inequality must be violated for sufficiently small $\rm{T_{1}}$. On the other hand, $\rm{f^{(1)}}$ is monotonically decreasing in $\rm{T_{1}}$ and tends to 0 for $\rm{T_{1}}\to \infty$. Thus, there will be a threshold value for $\rm{T_{1}}$, namely the one at which $\rm{f^{(1)}} = c(P)$, above which $\rm{W_{cyc} }<0$, and below which $\rm{W_{cyc} }\geq 0$$\;\;\square$\medskip

\emph{Remark}: We cannot give an analytical expression for the threshold $\rm{T_{1}^{0}}$, since finding it requires solving a transcendental equation. Note however that, since $\rm{c(P)}$ is monotonically increasing with $\rm{P}$,  $\rm{T_{1}^{0}}$, must become higher as $\rm{P}$ decreases. \medskip

\textbf{Result 2: [Conditions for `counter-rotating' heat engines]}  Suppose now $\rm{T_{1}<T_{2}}$. (i) For $\varepsilon_{3}/\varepsilon_{4}^{(1)}\leq1$ (weak coupling) $\rm{W_{cyc} }> 0$, i.e., a counter-rotating heat engine is not possible.  (ii) If $\varepsilon_{3}/\varepsilon_{4}^{(1)} > 1$  (strong coupling), then $\rm{W_{cyc}^{ad}} < 0$ for fixed $\rm{T_{2}}$ and sufficiently low $\rm{T_{1}  < T_{2}}$. In other words, in the adiabatic limit a counter-rotating heat engine is always possible, for sufficiently low $\rm{T_{1}}$. (iii) This conclusion continues to hold for any non-adiabatic cycle with $\rm{P \geq P_{\min}}$, in particular those with non-adiabatic strokes of sufficiently short duration $\tau$. 
\smallskip

\emph{Proof}: Define $x_{j} \coloneqq  \beta_{j}\varepsilon_{4}^{(j)}$, $y_{j} \coloneqq \varepsilon_{3}/\varepsilon_{4}^{(j)}$. For (i), note 
\begin{align}
\rm{f^{(1)}} &- \rm{f^{(2)}} =  g \left(x_{1}, y_{1}  \right) - g \left(x_{2}, y_{2}  \right) \nonumber \\
 &= \big[g \left(x_{1}, y_{1}  \right) - g \left(x_{2}, y_{1}  \right) \big]+\big[g \left(x_{2}, y_{1}  \right) - g \left(x_{2}, y_{2}  \right) \big].\nonumber
\end{align}
Since $\varepsilon_{4}^{(2)} \leq \varepsilon_{4}^{(1)}$, Eq.\,(\ref{eq:monoty}) implies the second term in square brackets is $>0$. Similarly, since also $\beta_{2} < \beta_{1}$, and $0\leq y_{1} \leq 1$,  Eq.\,(\ref{eq:monotx}) implies the first term is also $>0$. By Eq.\,(\ref{Wcyc}), this means $\rm{W_{cyc}^{ad}} > 0$. Recalling also that $\rm{W_{cyc}^{na}} > 0$, no heat engine is thus possible in this case. 

For (ii), note that, since now $y_{1} >1$, Eq.\,(\ref{eq:limitg}) implies that $\lim_{\beta_{1}\to\infty}\rm{f^{(1)}}$ = 0. Thus, for sufficiently high $\beta_{1}$ (i.e., low $\rm{T_{1}}$) we must have $\rm{f^{(1)} <  f^{(2)}}$. Referring again to Eq.\,(\ref{Wcyc}), this means $\rm{W_{cyc}^{ad}} < 0$.

Finally, for (iii), recall that the condition for a nonadiabatic heat engine is Eq.\,(\ref{nonadiabatic-cond}). Suppose $ \rm{c(P) > 0}$ (equivalently, $\rm{P > P_{\min}}$). Then, since  $\rm{f^{(1)}} \to 0$ as $\rm{T_{1}} \to 0$,  Eq.\,(\ref{nonadiabatic-cond}) must still be satisfied for sufficiently low $\rm{T_{1}}$.  
In particular, for very short nonadiabatic strokes  Eq.\,(\ref{Pinstant}) gives the limiting value $\rm{P(\tau \to 0) > P_{\min}}$. Indeed, replacing this above and simplifying, we find
\begin{equation}
\rm{c(P(\tau\to 0))} = \frac{h_{2}\varepsilon_{4}^{(1)}}{h_{1}\varepsilon_{4}^{(2)}} > 0.
\end{equation}
Since $\rm{c(P)}$ is continuous, Result 2(iii) must also hold for any sufficiently small $\tau \;\;\square$
\medskip

\textbf{Result 3 [Existence of a direct Temperature Gap]}: For $\varepsilon_{3}/\varepsilon_{4}^{(1)} > 1$  (strong coupling) there exists a range $\rm{(T_{1}^{a}, T_{1}^{b})}$ of values of $\rm{T_{1}}$  for which $\rm{W_{cyc} > 0}$ (i.e., a heat engine is not possible),  for any value of $\rm{T_{2}}$. 
Furthermore: given $\rm{h_{1}, h_{2}, J_{x}, J_{y}}$, there exists a temperature $\rm{T_{2}^{0}}$, valid for any $\rm{P > P_{\min}}$, such that
\begin{itemize}
\item[(i)] $\rm{T_{2}^{0}}$ is the value of $\rm{T_{2}}$ that maximizes the work $\rm{W_{2\rightarrow 1}}$ extracted during the expansion stroke,
\item[(ii)] $\rm{T_{2}^{0} \in(T_{1}^{a},T_{1}^{b})}$, and 
\item[(iii)] $\rm{W_{cyc}(T_{1},T_{2}) = 0}$ for $\rm{(T_{1}, T_{2}) = (T_{1}^{b} ,T_{2}^{0})}$ and $\rm{(T_{1}, T_{2}) = (T_{1}^{a} , T_{2}^{0})}$. In other words, the gap in $\rm{T_{1}}$ is `direct' and occurs at $\rm{T_{2}= T_{2}^{0}}$
\end{itemize}
\medskip

The proof of this result is entirely analogous to that of Result 2 in the Appendix of \cite{deoliveira2020efficiency}. This is possible because the properties of $g(x,y)$ listed in Eqs.\,(\ref{eq:gpositive})-(\ref{eq:gscaling}) above are the same as those in Eqs. (A4) - (A9) of \cite{deoliveira2020efficiency}. We omit the details, except to remark that the argument is essentially unaffected by the presence of nonadiabaticity, as long as $\rm{c(P) >0}$, i.e., $\rm{P>P_{\min}}$. The only relevant new observation is that the interval $\rm{(T_{1}^{a},T_{1}^{b})}$ widens monotonically as $\rm{P}$ decreases. This corresponds to the effect seen in Fig. \ref{j1_j2_}(d-f), whereby the upper engine zone shifts upwards, while the lower one is `squashed down'.

\bibliography{references}

\begin{thebibliography}{48}%
\makeatletter
\providecommand \@ifxundefined [1]{%
 \@ifx{#1\undefined}
}%
\providecommand \@ifnum [1]{%
 \ifnum #1\expandafter \@firstoftwo
 \else \expandafter \@secondoftwo
 \fi
}%
\providecommand \@ifx [1]{%
 \ifx #1\expandafter \@firstoftwo
 \else \expandafter \@secondoftwo
 \fi
}%
\providecommand \natexlab [1]{#1}%
\providecommand \enquote  [1]{``#1''}%
\providecommand \bibnamefont  [1]{#1}%
\providecommand \bibfnamefont [1]{#1}%
\providecommand \citenamefont [1]{#1}%
\providecommand \href@noop [0]{\@secondoftwo}%
\providecommand \href [0]{\begingroup \@sanitize@url \@href}%
\providecommand \@href[1]{\@@startlink{#1}\@@href}%
\providecommand \@@href[1]{\endgroup#1\@@endlink}%
\providecommand \@sanitize@url [0]{\catcode `\\12\catcode `\$12\catcode
  `\&12\catcode `\#12\catcode `\^12\catcode `\_12\catcode `\%12\relax}%
\providecommand \@@startlink[1]{}%
\providecommand \@@endlink[0]{}%
\providecommand \url  [0]{\begingroup\@sanitize@url \@url }%
\providecommand \@url [1]{\endgroup\@href {#1}{\urlprefix }}%
\providecommand \urlprefix  [0]{URL }%
\providecommand \Eprint [0]{\href }%
\providecommand \doibase [0]{https://doi.org/}%
\providecommand \selectlanguage [0]{\@gobble}%
\providecommand \bibinfo  [0]{\@secondoftwo}%
\providecommand \bibfield  [0]{\@secondoftwo}%
\providecommand \translation [1]{[#1]}%
\providecommand \BibitemOpen [0]{}%
\providecommand \bibitemStop [0]{}%
\providecommand \bibitemNoStop [0]{.\EOS\space}%
\providecommand \EOS [0]{\spacefactor3000\relax}%
\providecommand \BibitemShut  [1]{\csname bibitem#1\endcsname}%
\let\auto@bib@innerbib\@empty
\bibitem [{\citenamefont {Callen}(1985)}]{callen1985thermodynamics}%
  \BibitemOpen
  \bibfield  {author} {\bibinfo {author} {\bibfnamefont {H.~B.}\ \bibnamefont
  {Callen}},\ }\href@noop {} {\emph {\bibinfo {title} {Thermodynamics and an
  introduction to thermostatistics}}}\ (\bibinfo  {publisher} {Wiley},\
  \bibinfo {address} {New York},\ \bibinfo {year} {1985})\BibitemShut {NoStop}%
\bibitem [{\citenamefont {Kondepudi}\ and\ \citenamefont
  {Prigogine}(2014)}]{111837181X}%
  \BibitemOpen
  \bibfield  {author} {\bibinfo {author} {\bibfnamefont {D.}~\bibnamefont
  {Kondepudi}}\ and\ \bibinfo {author} {\bibfnamefont {I.}~\bibnamefont
  {Prigogine}},\ }\href@noop {} {\emph {\bibinfo {title} {Modern
  thermodynamics: from heat engines to dissipative structures}}}\ (\bibinfo
  {publisher} {Wiley},\ \bibinfo {address} {New York},\ \bibinfo {year}
  {2014})\BibitemShut {NoStop}%
\bibitem [{\citenamefont {Deffner}\ and\ \citenamefont
  {Campbell}(2019)}]{10.1088/2053-2571/ab21c6}%
  \BibitemOpen
  \bibfield  {author} {\bibinfo {author} {\bibfnamefont {S.}~\bibnamefont
  {Deffner}}\ and\ \bibinfo {author} {\bibfnamefont {S.}~\bibnamefont
  {Campbell}},\ }\href {https://doi.org/10.1088/2053-2571/ab21c6} {\emph
  {\bibinfo {title} {Quantum thermodynamics}}}\ (\bibinfo  {publisher} {Morgan
  \& Claypool Publishers},\ \bibinfo {address} {San Raphael, CA},\ \bibinfo
  {year} {2019})\BibitemShut {NoStop}%
\bibitem [{\citenamefont {Binder}\ \emph {et~al.}(2019)\citenamefont {Binder},
  \citenamefont {Correa}, \citenamefont {Gogolin}, \citenamefont {Anders},\
  and\ \citenamefont {Adesso}}]{binder2019thermodynamics}%
  \BibitemOpen
  \bibfield  {author} {\bibinfo {author} {\bibfnamefont {F.}~\bibnamefont
  {Binder}}, \bibinfo {author} {\bibfnamefont {L.}~\bibnamefont {Correa}},
  \bibinfo {author} {\bibfnamefont {C.}~\bibnamefont {Gogolin}}, \bibinfo
  {author} {\bibfnamefont {J.}~\bibnamefont {Anders}},\ and\ \bibinfo {author}
  {\bibfnamefont {G.}~\bibnamefont {Adesso}},\ }\href
  {https://books.google.com.br/books?id=5uWPDwAAQBAJ} {\emph {\bibinfo {title}
  {Thermodynamics in the Quantum Regime: Fundamental Aspects and New
  Directions}}},\ Fundamental Theories of Physics\ (\bibinfo  {publisher}
  {Springer International Publishing},\ \bibinfo {year} {2019})\BibitemShut
  {NoStop}%
\bibitem [{\citenamefont {Klaers}\ \emph {et~al.}(2017)\citenamefont {Klaers},
  \citenamefont {Faelt}, \citenamefont {Imamoglu},\ and\ \citenamefont
  {Togan}}]{PhysRevX.7.031044}%
  \BibitemOpen
  \bibfield  {author} {\bibinfo {author} {\bibfnamefont {J.}~\bibnamefont
  {Klaers}}, \bibinfo {author} {\bibfnamefont {S.}~\bibnamefont {Faelt}},
  \bibinfo {author} {\bibfnamefont {A.}~\bibnamefont {Imamoglu}},\ and\
  \bibinfo {author} {\bibfnamefont {E.}~\bibnamefont {Togan}},\ }\bibfield
  {title} {\bibinfo {title} {Squeezed thermal reservoirs as a resource for a
  nanomechanical engine beyond the carnot limit},\ }\href
  {https://doi.org/10.1103/PhysRevX.7.031044} {\bibfield  {journal} {\bibinfo
  {journal} {Physical Review X}\ }\textbf {\bibinfo {volume} {7}},\ \bibinfo
  {pages} {031044} (\bibinfo {year} {2017})}\BibitemShut {NoStop}%
\bibitem [{\citenamefont {Abah}\ and\ \citenamefont {Lutz}(2014)}]{Abah_2014}%
  \BibitemOpen
  \bibfield  {author} {\bibinfo {author} {\bibfnamefont {O.}~\bibnamefont
  {Abah}}\ and\ \bibinfo {author} {\bibfnamefont {E.}~\bibnamefont {Lutz}},\
  }\bibfield  {title} {\bibinfo {title} {Efficiency of heat engines coupled to
  nonequilibrium reservoirs},\ }\href
  {https://doi.org/10.1209/0295-5075/106/20001} {\bibfield  {journal} {\bibinfo
   {journal} {Europhysics Letters}\ }\textbf {\bibinfo {volume} {106}},\
  \bibinfo {pages} {20001} (\bibinfo {year} {2014})}\BibitemShut {NoStop}%
\bibitem [{\citenamefont {Huang}\ \emph {et~al.}(2012)\citenamefont {Huang},
  \citenamefont {Wang},\ and\ \citenamefont {Yi}}]{PhysRevE.86.051105}%
  \BibitemOpen
  \bibfield  {author} {\bibinfo {author} {\bibfnamefont {X.~L.}\ \bibnamefont
  {Huang}}, \bibinfo {author} {\bibfnamefont {T.}~\bibnamefont {Wang}},\ and\
  \bibinfo {author} {\bibfnamefont {X.~X.}\ \bibnamefont {Yi}},\ }\bibfield
  {title} {\bibinfo {title} {Effects of reservoir squeezing on quantum systems
  and work extraction},\ }\href {https://doi.org/10.1103/PhysRevE.86.051105}
  {\bibfield  {journal} {\bibinfo  {journal} {Physical Review E}\ }\textbf
  {\bibinfo {volume} {86}},\ \bibinfo {pages} {051105} (\bibinfo {year}
  {2012})}\BibitemShut {NoStop}%
\bibitem [{\citenamefont {Ro\ss{}nagel}\ \emph {et~al.}(2014)\citenamefont
  {Ro\ss{}nagel}, \citenamefont {Abah}, \citenamefont {Schmidt-Kaler},
  \citenamefont {Singer},\ and\ \citenamefont {Lutz}}]{PhysRevLett.112.030602}%
  \BibitemOpen
  \bibfield  {author} {\bibinfo {author} {\bibfnamefont {J.}~\bibnamefont
  {Ro\ss{}nagel}}, \bibinfo {author} {\bibfnamefont {O.}~\bibnamefont {Abah}},
  \bibinfo {author} {\bibfnamefont {F.}~\bibnamefont {Schmidt-Kaler}}, \bibinfo
  {author} {\bibfnamefont {K.}~\bibnamefont {Singer}},\ and\ \bibinfo {author}
  {\bibfnamefont {E.}~\bibnamefont {Lutz}},\ }\bibfield  {title} {\bibinfo
  {title} {Nanoscale heat engine beyond the carnot limit},\ }\href
  {https://doi.org/10.1103/PhysRevLett.112.030602} {\bibfield  {journal}
  {\bibinfo  {journal} {Physical Review Letters}\ }\textbf {\bibinfo {volume}
  {112}},\ \bibinfo {pages} {030602} (\bibinfo {year} {2014})}\BibitemShut
  {NoStop}%
\bibitem [{\citenamefont {Hardal}\ and\ \citenamefont {M{\"u}stecapl{\i}o{\u
  g}lu}(2015)}]{2015NatSR512953H}%
  \BibitemOpen
  \bibfield  {author} {\bibinfo {author} {\bibfnamefont {A.~{\"U}.~C.}\
  \bibnamefont {Hardal}}\ and\ \bibinfo {author} {\bibfnamefont {{\"O}.~E.}\
  \bibnamefont {M{\"u}stecapl{\i}o{\u g}lu}},\ }\bibfield  {title} {\bibinfo
  {title} {{Superradiant quantum heat engine}},\ }\href
  {https://doi.org/10.1038/srep12953} {\bibfield  {journal} {\bibinfo
  {journal} {Scientific Reports}\ }\textbf {\bibinfo {volume} {5}},\ \bibinfo
  {eid} {12953} (\bibinfo {year} {2015})}\BibitemShut {NoStop}%
\bibitem [{\citenamefont {Niedenzu}\ \emph {et~al.}(2016)\citenamefont
  {Niedenzu}, \citenamefont {Gelbwaser-Klimovsky}, \citenamefont {Kofman},\
  and\ \citenamefont {Kurizki}}]{Niedenzu2016}%
  \BibitemOpen
  \bibfield  {author} {\bibinfo {author} {\bibfnamefont {W.}~\bibnamefont
  {Niedenzu}}, \bibinfo {author} {\bibfnamefont {D.}~\bibnamefont
  {Gelbwaser-Klimovsky}}, \bibinfo {author} {\bibfnamefont {A.~G.}\
  \bibnamefont {Kofman}},\ and\ \bibinfo {author} {\bibfnamefont
  {G.}~\bibnamefont {Kurizki}},\ }\bibfield  {title} {\bibinfo {title} {On the
  operation of machines powered by quantum non-thermal baths},\ }\href
  {https://doi.org/10.1088/1367-2630/18/8/083012} {\bibfield  {journal}
  {\bibinfo  {journal} {New Journal of Physics}\ }\textbf {\bibinfo {volume}
  {18}},\ \bibinfo {pages} {083012} (\bibinfo {year} {2016})}\BibitemShut
  {NoStop}%
\bibitem [{\citenamefont {Manzano}\ \emph {et~al.}(2016)\citenamefont
  {Manzano}, \citenamefont {Galve}, \citenamefont {Zambrini},\ and\
  \citenamefont {Parrondo}}]{PhysRevE.93.052120}%
  \BibitemOpen
  \bibfield  {author} {\bibinfo {author} {\bibfnamefont {G.}~\bibnamefont
  {Manzano}}, \bibinfo {author} {\bibfnamefont {F.}~\bibnamefont {Galve}},
  \bibinfo {author} {\bibfnamefont {R.}~\bibnamefont {Zambrini}},\ and\
  \bibinfo {author} {\bibfnamefont {J.~M.~R.}\ \bibnamefont {Parrondo}},\
  }\bibfield  {title} {\bibinfo {title} {Entropy production and thermodynamic
  power of the squeezed thermal reservoir},\ }\href
  {https://doi.org/10.1103/PhysRevE.93.052120} {\bibfield  {journal} {\bibinfo
  {journal} {Physical Review E}\ }\textbf {\bibinfo {volume} {93}},\ \bibinfo
  {pages} {052120} (\bibinfo {year} {2016})}\BibitemShut {NoStop}%
\bibitem [{\citenamefont {Agarwalla}\ \emph {et~al.}(2017)\citenamefont
  {Agarwalla}, \citenamefont {Jiang},\ and\ \citenamefont
  {Segal}}]{PhysRevB.96.104304}%
  \BibitemOpen
  \bibfield  {author} {\bibinfo {author} {\bibfnamefont {B.~K.}\ \bibnamefont
  {Agarwalla}}, \bibinfo {author} {\bibfnamefont {J.-H.}\ \bibnamefont
  {Jiang}},\ and\ \bibinfo {author} {\bibfnamefont {D.}~\bibnamefont {Segal}},\
  }\bibfield  {title} {\bibinfo {title} {Quantum efficiency bound for
  continuous heat engines coupled to noncanonical reservoirs},\ }\href
  {https://doi.org/10.1103/PhysRevB.96.104304} {\bibfield  {journal} {\bibinfo
  {journal} {Physical Review B}\ }\textbf {\bibinfo {volume} {96}},\ \bibinfo
  {pages} {104304} (\bibinfo {year} {2017})}\BibitemShut {NoStop}%
\bibitem [{\citenamefont {Stefanatos}(2014)}]{PhysRevE.90.012119}%
  \BibitemOpen
  \bibfield  {author} {\bibinfo {author} {\bibfnamefont {D.}~\bibnamefont
  {Stefanatos}},\ }\bibfield  {title} {\bibinfo {title} {Optimal efficiency of
  a noisy quantum heat engine},\ }\href
  {https://doi.org/10.1103/PhysRevE.90.012119} {\bibfield  {journal} {\bibinfo
  {journal} {Physical Review E}\ }\textbf {\bibinfo {volume} {90}},\ \bibinfo
  {pages} {012119} (\bibinfo {year} {2014})}\BibitemShut {NoStop}%
\bibitem [{\citenamefont {Torrontegui}\ and\ \citenamefont
  {Kosloff}(2013)}]{PhysRevE.88.032103}%
  \BibitemOpen
  \bibfield  {author} {\bibinfo {author} {\bibfnamefont {E.}~\bibnamefont
  {Torrontegui}}\ and\ \bibinfo {author} {\bibfnamefont {R.}~\bibnamefont
  {Kosloff}},\ }\bibfield  {title} {\bibinfo {title} {Quest for absolute zero
  in the presence of external noise},\ }\href
  {https://doi.org/10.1103/PhysRevE.88.032103} {\bibfield  {journal} {\bibinfo
  {journal} {Physical Review E}\ }\textbf {\bibinfo {volume} {88}},\ \bibinfo
  {pages} {032103} (\bibinfo {year} {2013})}\BibitemShut {NoStop}%
\bibitem [{\citenamefont {de~Oliveira}\ and\ \citenamefont
  {Jonathan}(2021)}]{deoliveira2020efficiency}%
  \BibitemOpen
  \bibfield  {author} {\bibinfo {author} {\bibfnamefont {T.~R.}\ \bibnamefont
  {de~Oliveira}}\ and\ \bibinfo {author} {\bibfnamefont {D.}~\bibnamefont
  {Jonathan}},\ }\bibfield  {title} {\bibinfo {title} {Efficiency gain and
  bidirectional operation of quantum engines with decoupled internal levels},\
  }\href {https://doi.org/https://doi.org/10.1103/PhysRevE.104.044133}
  {\bibfield  {journal} {\bibinfo  {journal} {Phys. Rev. E}\ }\textbf {\bibinfo
  {volume} {104}},\ \bibinfo {pages} {044133} (\bibinfo {year}
  {2021})}\BibitemShut {NoStop}%
\bibitem [{\citenamefont {Tajima}\ and\ \citenamefont
  {Funo}(2021)}]{PhysRevLett.127.190604}%
  \BibitemOpen
  \bibfield  {author} {\bibinfo {author} {\bibfnamefont {H.}~\bibnamefont
  {Tajima}}\ and\ \bibinfo {author} {\bibfnamefont {K.}~\bibnamefont {Funo}},\
  }\bibfield  {title} {\bibinfo {title} {Superconducting-like heat current:
  Effective cancellation of current-dissipation trade-off by quantum
  coherence},\ }\href {https://doi.org/10.1103/PhysRevLett.127.190604}
  {\bibfield  {journal} {\bibinfo  {journal} {Phys. Rev. Lett.}\ }\textbf
  {\bibinfo {volume} {127}},\ \bibinfo {pages} {190604} (\bibinfo {year}
  {2021})}\BibitemShut {NoStop}%
\bibitem [{\citenamefont {Geva}\ and\ \citenamefont
  {Kosloff}(1992)}]{doi:10.1063/1.461951}%
  \BibitemOpen
  \bibfield  {author} {\bibinfo {author} {\bibfnamefont {E.}~\bibnamefont
  {Geva}}\ and\ \bibinfo {author} {\bibfnamefont {R.}~\bibnamefont {Kosloff}},\
  }\bibfield  {title} {\bibinfo {title} {A quantum‐mechanical heat engine
  operating in finite time. a model consisting of spin‐1/2 systems as the
  working fluid},\ }\href {https://doi.org/10.1063/1.461951} {\bibfield
  {journal} {\bibinfo  {journal} {The Journal of Chemical Physics}\ }\textbf
  {\bibinfo {volume} {96}},\ \bibinfo {pages} {3054} (\bibinfo {year}
  {1992})},\ \Eprint {https://arxiv.org/abs/https://doi.org/10.1063/1.461951}
  {https://doi.org/10.1063/1.461951} \BibitemShut {NoStop}%
\bibitem [{\citenamefont {Kieu}(2004)}]{PhysRevLett.93.140403}%
  \BibitemOpen
  \bibfield  {author} {\bibinfo {author} {\bibfnamefont {T.~D.}\ \bibnamefont
  {Kieu}},\ }\bibfield  {title} {\bibinfo {title} {The second law, maxwell's
  demon, and work derivable from quantum heat engines},\ }\href
  {https://doi.org/10.1103/PhysRevLett.93.140403} {\bibfield  {journal}
  {\bibinfo  {journal} {Physical Review Letters}\ }\textbf {\bibinfo {volume}
  {93}},\ \bibinfo {pages} {140403} (\bibinfo {year} {2004})}\BibitemShut
  {NoStop}%
\bibitem [{\citenamefont {Quan}\ \emph {et~al.}(2007)\citenamefont {Quan},
  \citenamefont {Liu}, \citenamefont {Sun},\ and\ \citenamefont
  {Nori}}]{PhysRevE.76.031105}%
  \BibitemOpen
  \bibfield  {author} {\bibinfo {author} {\bibfnamefont {H.~T.}\ \bibnamefont
  {Quan}}, \bibinfo {author} {\bibfnamefont {Y.-X.}\ \bibnamefont {Liu}},
  \bibinfo {author} {\bibfnamefont {C.~P.}\ \bibnamefont {Sun}},\ and\ \bibinfo
  {author} {\bibfnamefont {F.}~\bibnamefont {Nori}},\ }\bibfield  {title}
  {\bibinfo {title} {Quantum thermodynamic cycles and quantum heat engines},\
  }\href {https://doi.org/10.1103/PhysRevE.76.031105} {\bibfield  {journal}
  {\bibinfo  {journal} {Physical Review E}\ }\textbf {\bibinfo {volume} {76}},\
  \bibinfo {pages} {031105} (\bibinfo {year} {2007})}\BibitemShut {NoStop}%
\bibitem [{\citenamefont {Thomas}\ and\ \citenamefont
  {Johal}(2011)}]{PhysRevE.83.031135}%
  \BibitemOpen
  \bibfield  {author} {\bibinfo {author} {\bibfnamefont {G.}~\bibnamefont
  {Thomas}}\ and\ \bibinfo {author} {\bibfnamefont {R.~S.}\ \bibnamefont
  {Johal}},\ }\bibfield  {title} {\bibinfo {title} {Coupled quantum otto
  cycle},\ }\href {https://doi.org/10.1103/PhysRevE.83.031135} {\bibfield
  {journal} {\bibinfo  {journal} {Phys. Rev. E}\ }\textbf {\bibinfo {volume}
  {83}},\ \bibinfo {pages} {031135} (\bibinfo {year} {2011})}\BibitemShut
  {NoStop}%
\bibitem [{\citenamefont {Quan}\ \emph {et~al.}(2005)\citenamefont {Quan},
  \citenamefont {Zhang},\ and\ \citenamefont {Sun}}]{PhysRevE.72.056110}%
  \BibitemOpen
  \bibfield  {author} {\bibinfo {author} {\bibfnamefont {H.~T.}\ \bibnamefont
  {Quan}}, \bibinfo {author} {\bibfnamefont {P.}~\bibnamefont {Zhang}},\ and\
  \bibinfo {author} {\bibfnamefont {C.~P.}\ \bibnamefont {Sun}},\ }\bibfield
  {title} {\bibinfo {title} {Quantum heat engine with multilevel quantum
  systems},\ }\href {https://doi.org/10.1103/PhysRevE.72.056110} {\bibfield
  {journal} {\bibinfo  {journal} {Physical Review E}\ }\textbf {\bibinfo
  {volume} {72}},\ \bibinfo {pages} {056110} (\bibinfo {year}
  {2005})}\BibitemShut {NoStop}%
\bibitem [{\citenamefont {Solfanelli}\ \emph {et~al.}(2020)\citenamefont
  {Solfanelli}, \citenamefont {Falsetti},\ and\ \citenamefont
  {Campisi}}]{PhysRevB.101.054513}%
  \BibitemOpen
  \bibfield  {author} {\bibinfo {author} {\bibfnamefont {A.}~\bibnamefont
  {Solfanelli}}, \bibinfo {author} {\bibfnamefont {M.}~\bibnamefont
  {Falsetti}},\ and\ \bibinfo {author} {\bibfnamefont {M.}~\bibnamefont
  {Campisi}},\ }\bibfield  {title} {\bibinfo {title} {Nonadiabatic single-qubit
  quantum otto engine},\ }\href {https://doi.org/10.1103/PhysRevB.101.054513}
  {\bibfield  {journal} {\bibinfo  {journal} {Phys. Rev. B}\ }\textbf {\bibinfo
  {volume} {101}},\ \bibinfo {pages} {054513} (\bibinfo {year}
  {2020})}\BibitemShut {NoStop}%
\bibitem [{\citenamefont {Karimi}\ and\ \citenamefont
  {Pekola}(2016)}]{PhysRevB.94.184503}%
  \BibitemOpen
  \bibfield  {author} {\bibinfo {author} {\bibfnamefont {B.}~\bibnamefont
  {Karimi}}\ and\ \bibinfo {author} {\bibfnamefont {J.~P.}\ \bibnamefont
  {Pekola}},\ }\bibfield  {title} {\bibinfo {title} {Otto refrigerator based on
  a superconducting qubit: Classical and quantum performance},\ }\href
  {https://doi.org/10.1103/PhysRevB.94.184503} {\bibfield  {journal} {\bibinfo
  {journal} {Phys. Rev. B}\ }\textbf {\bibinfo {volume} {94}},\ \bibinfo
  {pages} {184503} (\bibinfo {year} {2016})}\BibitemShut {NoStop}%
\bibitem [{\citenamefont {{\c{C}}akmak}\ \emph {et~al.}(2016)\citenamefont
  {{\c{C}}akmak}, \citenamefont {Altintas},\ and\ \citenamefont
  {E.~M{\"u}stecapl{\i}o{\u{g}}lu}}]{Cakmak2016}%
  \BibitemOpen
  \bibfield  {author} {\bibinfo {author} {\bibfnamefont {S.}~\bibnamefont
  {{\c{C}}akmak}}, \bibinfo {author} {\bibfnamefont {F.}~\bibnamefont
  {Altintas}},\ and\ \bibinfo {author} {\bibfnamefont {{\"O}.}~\bibnamefont
  {E.~M{\"u}stecapl{\i}o{\u{g}}lu}},\ }\bibfield  {title} {\bibinfo {title}
  {Lipkin-meshkov-glick model in a quantum otto cycle},\ }\href
  {https://doi.org/10.1140/epjp/i2016-16197-0} {\bibfield  {journal} {\bibinfo
  {journal} {The European Physical Journal Plus}\ }\textbf {\bibinfo {volume}
  {131}},\ \bibinfo {pages} {197} (\bibinfo {year} {2016})}\BibitemShut
  {NoStop}%
\bibitem [{\citenamefont {{\c{C}}akmak}\ and\ \citenamefont
  {M{\"u}stecapl{\i}o{\u{g}}lu}(2019)}]{PhysRevE.99.032108}%
  \BibitemOpen
  \bibfield  {author} {\bibinfo {author} {\bibfnamefont {B.}~\bibnamefont
  {{\c{C}}akmak}}\ and\ \bibinfo {author} {\bibfnamefont {{\"O}.~E.}\
  \bibnamefont {M{\"u}stecapl{\i}o{\u{g}}lu}},\ }\bibfield  {title} {\bibinfo
  {title} {Spin quantum heat engines with shortcuts to adiabaticity},\ }\href
  {https://doi.org/10.1103/PhysRevE.99.032108} {\bibfield  {journal} {\bibinfo
  {journal} {Phys. Rev. E}\ }\textbf {\bibinfo {volume} {99}},\ \bibinfo
  {pages} {032108} (\bibinfo {year} {2019})}\BibitemShut {NoStop}%
\bibitem [{Note1()}]{Note1}%
  \BibitemOpen
  \bibinfo {note} {Note that, throughout this article, we use units such that
  $\hbar $, $k_B$ and the magnetic moment $\mu $ of each spin are all equal to
  1}\BibitemShut {NoStop}%
\bibitem [{\citenamefont {Alicki}(1979)}]{0305-4470-12-5-007}%
  \BibitemOpen
  \bibfield  {author} {\bibinfo {author} {\bibfnamefont {R.}~\bibnamefont
  {Alicki}},\ }\bibfield  {title} {\bibinfo {title} {The quantum open system as
  a model of the heat engine},\ }\href
  {https://doi.org/10.1088/0305-4470/12/5/007} {\bibfield  {journal} {\bibinfo
  {journal} {Journal of Physics A: \textnormal{mathematical and general}}\
  }\textbf {\bibinfo {volume} {12}},\ \bibinfo {pages} {L103} (\bibinfo {year}
  {1979})}\BibitemShut {NoStop}%
\bibitem [{\citenamefont {Kosloff}(1984)}]{doi:10.1063/1.446862}%
  \BibitemOpen
  \bibfield  {author} {\bibinfo {author} {\bibfnamefont {R.}~\bibnamefont
  {Kosloff}},\ }\bibfield  {title} {\bibinfo {title} {A quantum mechanical open
  system as a model of a heat engine},\ }\href
  {https://doi.org/10.1063/1.446862} {\bibfield  {journal} {\bibinfo  {journal}
  {The Journal of Chemical Physics}\ }\textbf {\bibinfo {volume} {80}},\
  \bibinfo {pages} {1625} (\bibinfo {year} {1984})},\ \Eprint
  {https://arxiv.org/abs/https://doi.org/10.1063/1.446862}
  {https://doi.org/10.1063/1.446862} \BibitemShut {NoStop}%
\bibitem [{\citenamefont {Allahverdyan}\ \emph {et~al.}(2004)\citenamefont
  {Allahverdyan}, \citenamefont {Balian},\ and\ \citenamefont
  {Nieuwenhuizen}}]{Allahverdyan_2004}%
  \BibitemOpen
  \bibfield  {author} {\bibinfo {author} {\bibfnamefont {A.~E.}\ \bibnamefont
  {Allahverdyan}}, \bibinfo {author} {\bibfnamefont {R.}~\bibnamefont
  {Balian}},\ and\ \bibinfo {author} {\bibfnamefont {T.~M.}\ \bibnamefont
  {Nieuwenhuizen}},\ }\bibfield  {title} {\bibinfo {title} {Maximal work
  extraction from finite quantum systems},\ }\href
  {https://doi.org/10.1209/epl/i2004-10101-2} {\bibfield  {journal} {\bibinfo
  {journal} {Europhysics Letters}\ }\textbf {\bibinfo {volume} {67}},\ \bibinfo
  {pages} {565} (\bibinfo {year} {2004})}\BibitemShut {NoStop}%
\bibitem [{\citenamefont {Campisi}\ \emph {et~al.}(2011)\citenamefont
  {Campisi}, \citenamefont {H\"anggi},\ and\ \citenamefont
  {Talkner}}]{RevModPhys.83.771}%
  \BibitemOpen
  \bibfield  {author} {\bibinfo {author} {\bibfnamefont {M.}~\bibnamefont
  {Campisi}}, \bibinfo {author} {\bibfnamefont {P.}~\bibnamefont {H\"anggi}},\
  and\ \bibinfo {author} {\bibfnamefont {P.}~\bibnamefont {Talkner}},\
  }\bibfield  {title} {\bibinfo {title} {Colloquium: quantum fluctuation
  relations: foundations and applications},\ }\href
  {https://doi.org/10.1103/RevModPhys.83.771} {\bibfield  {journal} {\bibinfo
  {journal} {Reviews of Modern Physics}\ }\textbf {\bibinfo {volume} {83}},\
  \bibinfo {pages} {771} (\bibinfo {year} {2011})}\BibitemShut {NoStop}%
\bibitem [{\citenamefont {Niedenzu}\ \emph {et~al.}(2018)\citenamefont
  {Niedenzu}, \citenamefont {Mukherjee}, \citenamefont {Ghosh}, \citenamefont
  {Kofman},\ and\ \citenamefont {Kurizki}}]{Niedenzu2018}%
  \BibitemOpen
  \bibfield  {author} {\bibinfo {author} {\bibfnamefont {W.}~\bibnamefont
  {Niedenzu}}, \bibinfo {author} {\bibfnamefont {V.}~\bibnamefont {Mukherjee}},
  \bibinfo {author} {\bibfnamefont {A.}~\bibnamefont {Ghosh}}, \bibinfo
  {author} {\bibfnamefont {A.~G.}\ \bibnamefont {Kofman}},\ and\ \bibinfo
  {author} {\bibfnamefont {G.}~\bibnamefont {Kurizki}},\ }\bibfield  {title}
  {\bibinfo {title} {{Quantum engine efficiency bound beyond the second law of
  thermodynamics}},\ }\href {https://doi.org/10.1038/s41467-017-01991-6}
  {\bibfield  {journal} {\bibinfo  {journal} {Nature Communications}\ }\textbf
  {\bibinfo {volume} {9}},\ \bibinfo {eid} {165} (\bibinfo {year}
  {2018})}\BibitemShut {NoStop}%
\bibitem [{\citenamefont {Bernardo}(2020)}]{PhysRevE.102.062152}%
  \BibitemOpen
  \bibfield  {author} {\bibinfo {author} {\bibfnamefont {B.~d.~L.}\
  \bibnamefont {Bernardo}},\ }\bibfield  {title} {\bibinfo {title} {Unraveling
  the role of coherence in the first law of quantum thermodynamics},\ }\href
  {https://doi.org/10.1103/PhysRevE.102.062152} {\bibfield  {journal} {\bibinfo
   {journal} {Phys. Rev. E}\ }\textbf {\bibinfo {volume} {102}},\ \bibinfo
  {pages} {062152} (\bibinfo {year} {2020})}\BibitemShut {NoStop}%
\bibitem [{\citenamefont {{Alipour}}\ \emph {et~al.}(2019)\citenamefont
  {{Alipour}}, \citenamefont {{Rezakhani}}, \citenamefont {{Chenu}},
  \citenamefont {{del Campo}},\ and\ \citenamefont
  {{Ala-Nissila}}}]{2019arXiv191201939A}%
  \BibitemOpen
  \bibfield  {author} {\bibinfo {author} {\bibfnamefont {S.}~\bibnamefont
  {{Alipour}}}, \bibinfo {author} {\bibfnamefont {A.~T.}\ \bibnamefont
  {{Rezakhani}}}, \bibinfo {author} {\bibfnamefont {A.}~\bibnamefont
  {{Chenu}}}, \bibinfo {author} {\bibfnamefont {A.}~\bibnamefont {{del
  Campo}}},\ and\ \bibinfo {author} {\bibfnamefont {T.}~\bibnamefont
  {{Ala-Nissila}}},\ }\bibfield  {title} {\bibinfo {title} {{Entropy-Based
  Formulation of Thermodynamics in Arbitrary Quantum Evolution}},\ }\href@noop
  {} {\bibfield  {journal} {\bibinfo  {journal} {arXiv e-prints}\ ,\ \bibinfo
  {eid} {arXiv:1912.01939}} (\bibinfo {year} {2019})},\ \Eprint
  {https://arxiv.org/abs/1912.01939} {arXiv:1912.01939 [quant-ph]} \BibitemShut
  {NoStop}%
\bibitem [{\citenamefont {Su}\ \emph {et~al.}(2018)\citenamefont {Su},
  \citenamefont {Chen}, \citenamefont {Ma}, \citenamefont {Chen},\ and\
  \citenamefont {Sun}}]{Su_2018}%
  \BibitemOpen
  \bibfield  {author} {\bibinfo {author} {\bibfnamefont {S.}~\bibnamefont
  {Su}}, \bibinfo {author} {\bibfnamefont {J.}~\bibnamefont {Chen}}, \bibinfo
  {author} {\bibfnamefont {Y.}~\bibnamefont {Ma}}, \bibinfo {author}
  {\bibfnamefont {J.}~\bibnamefont {Chen}},\ and\ \bibinfo {author}
  {\bibfnamefont {C.}~\bibnamefont {Sun}},\ }\bibfield  {title} {\bibinfo
  {title} {The heat and work of quantum thermodynamic processes with quantum
  coherence},\ }\href {https://doi.org/10.1088/1674-1056/27/6/060502}
  {\bibfield  {journal} {\bibinfo  {journal} {Chinese Physics B}\ }\textbf
  {\bibinfo {volume} {27}},\ \bibinfo {pages} {060502} (\bibinfo {year}
  {2018})}\BibitemShut {NoStop}%
\bibitem [{\citenamefont {{Ahmadi}}\ \emph {et~al.}(2019)\citenamefont
  {{Ahmadi}}, \citenamefont {{Salimi}},\ and\ \citenamefont
  {{Khorashad}}}]{2019arXiv191201983A}%
  \BibitemOpen
  \bibfield  {author} {\bibinfo {author} {\bibfnamefont {B.}~\bibnamefont
  {{Ahmadi}}}, \bibinfo {author} {\bibfnamefont {S.}~\bibnamefont {{Salimi}}},\
  and\ \bibinfo {author} {\bibfnamefont {A.~S.}\ \bibnamefont {{Khorashad}}},\
  }\bibfield  {title} {\bibinfo {title} {{Refined Definitions of Heat and Work
  in Quantum Thermodynamics}},\ }\href@noop {} {\bibfield  {journal} {\bibinfo
  {journal} {arXiv e-prints}\ } (\bibinfo {year} {2019})},\ \Eprint
  {https://arxiv.org/abs/1912.01983} {arXiv:1912.01983 [quant-ph]} \BibitemShut
  {NoStop}%
\bibitem [{\citenamefont {Messiah}(1962)}]{Messiah62}%
  \BibitemOpen
  \bibfield  {author} {\bibinfo {author} {\bibfnamefont {A.}~\bibnamefont
  {Messiah}},\ }\href@noop {} {\emph {\bibinfo {title} {Quantum Mechanics}}},\
  Vol.~\bibinfo {volume} {II}\ (\bibinfo  {publisher} {North-Holland},\
  \bibinfo {year} {1962})\BibitemShut {NoStop}%
\bibitem [{\citenamefont {Rezek}\ and\ \citenamefont
  {Kosloff}(2006)}]{Rezek_2006}%
  \BibitemOpen
  \bibfield  {author} {\bibinfo {author} {\bibfnamefont {Y.}~\bibnamefont
  {Rezek}}\ and\ \bibinfo {author} {\bibfnamefont {R.}~\bibnamefont
  {Kosloff}},\ }\bibfield  {title} {\bibinfo {title} {Irreversible performance
  of a quantum harmonic heat engine},\ }\href
  {https://doi.org/10.1088/1367-2630/8/5/083} {\bibfield  {journal} {\bibinfo
  {journal} {New Journal of Physics}\ }\textbf {\bibinfo {volume} {8}},\
  \bibinfo {pages} {83} (\bibinfo {year} {2006})}\BibitemShut {NoStop}%
\bibitem [{\citenamefont {Rezek}(2010)}]{e12081885}%
  \BibitemOpen
  \bibfield  {author} {\bibinfo {author} {\bibfnamefont {Y.}~\bibnamefont
  {Rezek}},\ }\bibfield  {title} {\bibinfo {title} {Reflections on friction in
  quantum mechanics},\ }\href {https://doi.org/10.3390/e12081885} {\bibfield
  {journal} {\bibinfo  {journal} {Entropy}\ }\textbf {\bibinfo {volume} {12}},\
  \bibinfo {pages} {1885} (\bibinfo {year} {2010})}\BibitemShut {NoStop}%
\bibitem [{\citenamefont {Rigolin}\ \emph {et~al.}(2008)\citenamefont
  {Rigolin}, \citenamefont {Ortiz},\ and\ \citenamefont
  {Ponce}}]{PhysRevA.78.052508}%
  \BibitemOpen
  \bibfield  {author} {\bibinfo {author} {\bibfnamefont {G.}~\bibnamefont
  {Rigolin}}, \bibinfo {author} {\bibfnamefont {G.}~\bibnamefont {Ortiz}},\
  and\ \bibinfo {author} {\bibfnamefont {V.~H.}\ \bibnamefont {Ponce}},\
  }\bibfield  {title} {\bibinfo {title} {Beyond the quantum adiabatic
  approximation: Adiabatic perturbation theory},\ }\href
  {https://doi.org/10.1103/PhysRevA.78.052508} {\bibfield  {journal} {\bibinfo
  {journal} {Phys. Rev. A}\ }\textbf {\bibinfo {volume} {78}},\ \bibinfo
  {pages} {052508} (\bibinfo {year} {2008})}\BibitemShut {NoStop}%
\bibitem [{\citenamefont {Chen}\ \emph {et~al.}(2019)\citenamefont {Chen},
  \citenamefont {Sun},\ and\ \citenamefont {Dong}}]{PhysRevE.100.032144}%
  \BibitemOpen
  \bibfield  {author} {\bibinfo {author} {\bibfnamefont {J.-F.}\ \bibnamefont
  {Chen}}, \bibinfo {author} {\bibfnamefont {C.-P.}\ \bibnamefont {Sun}},\ and\
  \bibinfo {author} {\bibfnamefont {H.}~\bibnamefont {Dong}},\ }\bibfield
  {title} {\bibinfo {title} {Boosting the performance of quantum otto heat
  engines},\ }\href {https://doi.org/10.1103/PhysRevE.100.032144} {\bibfield
  {journal} {\bibinfo  {journal} {Phys. Rev. E}\ }\textbf {\bibinfo {volume}
  {100}},\ \bibinfo {pages} {032144} (\bibinfo {year} {2019})}\BibitemShut
  {NoStop}%
\bibitem [{\citenamefont {Teufel}(2003)}]{teufel2003adiabatic}%
  \BibitemOpen
  \bibfield  {author} {\bibinfo {author} {\bibfnamefont {S.}~\bibnamefont
  {Teufel}},\ }\href@noop {} {\emph {\bibinfo {title} {Adiabatic Perturbation
  Theory in Quantum Dynamics}}},\ Lecture notes in mathematics\ (\bibinfo
  {publisher} {Springer},\ \bibinfo {year} {2003})\BibitemShut {NoStop}%
\bibitem [{\citenamefont {Buffoni}\ \emph {et~al.}(2019)\citenamefont
  {Buffoni}, \citenamefont {Solfanelli}, \citenamefont {Verrucchi},
  \citenamefont {Cuccoli},\ and\ \citenamefont {Campisi}}]{Buffoni19}%
  \BibitemOpen
  \bibfield  {author} {\bibinfo {author} {\bibfnamefont {L.}~\bibnamefont
  {Buffoni}}, \bibinfo {author} {\bibfnamefont {A.}~\bibnamefont {Solfanelli}},
  \bibinfo {author} {\bibfnamefont {P.}~\bibnamefont {Verrucchi}}, \bibinfo
  {author} {\bibfnamefont {A.}~\bibnamefont {Cuccoli}},\ and\ \bibinfo {author}
  {\bibfnamefont {M.}~\bibnamefont {Campisi}},\ }\bibfield  {title} {\bibinfo
  {title} {Quantum measurement cooling},\ }\href
  {https://doi.org/10.1103/PhysRevLett.122.070603} {\bibfield  {journal}
  {\bibinfo  {journal} {Phys. Rev. Lett.}\ }\textbf {\bibinfo {volume} {122}},\
  \bibinfo {pages} {070603} (\bibinfo {year} {2019})}\BibitemShut {NoStop}%
\bibitem [{Note2()}]{Note2}%
  \BibitemOpen
  \bibinfo {note} {In Appendix \ref {Formal-analysis} we show it is in fact
  also a necessary condition}\BibitemShut {NoStop}%
\bibitem [{Note3()}]{Note3}%
  \BibitemOpen
  \bibinfo {note} {Recall that, as seen in the last section, no engine can
  exist if $\protect \rm {P}<0.5$}\BibitemShut {NoStop}%
\bibitem [{\citenamefont {Kosloff}\ and\ \citenamefont
  {Rezek}(2017)}]{e19040136}%
  \BibitemOpen
  \bibfield  {author} {\bibinfo {author} {\bibfnamefont {R.}~\bibnamefont
  {Kosloff}}\ and\ \bibinfo {author} {\bibfnamefont {Y.}~\bibnamefont
  {Rezek}},\ }\bibfield  {title} {\bibinfo {title} {The quantum harmonic otto
  cycle},\ }\bibfield  {journal} {\bibinfo  {journal} {Entropy}\ }\textbf
  {\bibinfo {volume} {19}},\ \href {https://doi.org/10.3390/e19040136}
  {10.3390/e19040136} (\bibinfo {year} {2017})\BibitemShut {NoStop}%
\bibitem [{\citenamefont {Kosloff}\ and\ \citenamefont
  {Feldmann}(2002)}]{PhysRevE.65.055102}%
  \BibitemOpen
  \bibfield  {author} {\bibinfo {author} {\bibfnamefont {R.}~\bibnamefont
  {Kosloff}}\ and\ \bibinfo {author} {\bibfnamefont {T.}~\bibnamefont
  {Feldmann}},\ }\bibfield  {title} {\bibinfo {title} {Discrete four-stroke
  quantum heat engine exploring the origin of friction},\ }\href
  {https://doi.org/10.1103/PhysRevE.65.055102} {\bibfield  {journal} {\bibinfo
  {journal} {Phys. Rev. E}\ }\textbf {\bibinfo {volume} {65}},\ \bibinfo
  {pages} {055102} (\bibinfo {year} {2002})}\BibitemShut {NoStop}%
\bibitem [{\citenamefont {Anka}\ \emph {et~al.}(2021)\citenamefont {Anka},
  \citenamefont {de~Oliveira},\ and\ \citenamefont
  {Jonathan}}]{PhysRevE.104.054128}%
  \BibitemOpen
  \bibfield  {author} {\bibinfo {author} {\bibfnamefont {M.~F.}\ \bibnamefont
  {Anka}}, \bibinfo {author} {\bibfnamefont {T.~R.}\ \bibnamefont
  {de~Oliveira}},\ and\ \bibinfo {author} {\bibfnamefont {D.}~\bibnamefont
  {Jonathan}},\ }\bibfield  {title} {\bibinfo {title} {Measurement-based
  quantum heat engine in a multilevel system},\ }\href
  {https://doi.org/10.1103/PhysRevE.104.054128} {\bibfield  {journal} {\bibinfo
   {journal} {Phys. Rev. E}\ }\textbf {\bibinfo {volume} {104}},\ \bibinfo
  {pages} {054128} (\bibinfo {year} {2021})}\BibitemShut {NoStop}%
\bibitem [{\citenamefont {{\c C}akmak}(2021)}]{bla}%
  \BibitemOpen
  \bibfield  {author} {\bibinfo {author} {\bibfnamefont {B.}~\bibnamefont {{\c
  C}akmak}},\ }\bibfield  {title} {\bibinfo {title} {Finite-time two-spin
  quantum otto engines: Shortcuts to adiabaticity vs. irreversibility},\ }\href
  {https://doi.org/10.3906/fiz-2101-10} {\bibfield  {journal} {\bibinfo
  {journal} {Turkish Journal of Physics}\ }\textbf {\bibinfo {volume} {45}},\
  \bibinfo {pages} {59} (\bibinfo {year} {2021})}\BibitemShut {NoStop}%
\end{thebibliography}%

\end{document}